%% file: main.tex
\documentclass[ALICE,manyauthors]{cernphprep}
\usepackage[comma,square,numbers,sort&compress]{natbib}
\usepackage[hypertexnames=false]{hyperref}
\usepackage{lineno}
\usepackage{xspace}
\usepackage{comment}
\usepackage[T1]{fontenc} 
\usepackage{orcidlink}
\usepackage{float}
\usepackage{units}
\usepackage{booktabs}
\usepackage{xcolor}

\begin{document}
\input{commands.tex}

\begin{titlepage}
    \PHyear{2026}       
    \PHnumber{281}      
    \PHdate{8 October}  

    \title{Evidence of partonic collectivity in OO collisions through the measurement of $\phi$, \kzero, and $\Lambda$ elliptic flow}
    \ShortTitle{Partonic flow in OO collisions}   

    \Collaboration{ALICE Collaboration\thanks{See Appendix~\ref{app:collab} for the list of collaboration members}}
    \ShortAuthor{ALICE Collaboration} 

    \begin{abstract}
        The first measurement of the elliptic-flow coefficient ($v_{2}$) of $\phi$, \kzero, and $\Lambda (\overline{\Lambda})$ hadrons performed in OO collisions at $\sqrt{s_{\mathrm{NN}}}=5.36$~TeV with the ALICE detector is presented as a function of transverse momentum (\pt) in central and semi-central collisions. For the $\phi$ meson, in particular, this is the first measurement of the elliptic flow in a small collision system. At low \pt, a characteristic mass ordering of \vtwo\ is observed, consistent with collective expansion of the system, in which heavier hadrons are shifted to higher \pt by a common velocity field. At intermediate \pt, above \unit[3]{\GeVc}, a baryon--meson splitting with a $10\sigma$ significance emerges, with the $\Lambda (\overline{\Lambda})$ baryon exhibiting larger \vtwo\ values than mesons, while, despite its mass being close to the mass of the $\Lambda$, the $\phi$ follows the K$^{0}_{\rm S}$. The $\phi$ meson behavior disfavors a purely mass-driven interpretation of the observed patterns and supports the role of hadronization via quark recombination from a collectively flowing partonic medium. These results provide complementary evidence that key signatures of partonic collectivity observed in \PbPb collisions persist in light nucleus--nucleus systems.
    \end{abstract}
\end{titlepage}

\setcounter{page}{2} 


Quantum chromodynamics (QCD) predicts the emergence of a deconfined state of strongly-interacting matter, the quark–gluon plasma (QGP), at the extreme energy densities achieved in ultrarelativistic heavy-ion collisions at the Super Proton Synchrotron (SPS), the Relativistic Heavy Ion Collider (RHIC), and the Large Hadron Collider (LHC)~\cite{Shuryak:1980tp, Cleymans:1985wb, Schukraft:1993fcq, Bass:1998vz, Heinz:2008tv, Busza:2018rrf, STAR:2005gfr, PHENIX:2004vcz, BRAHMS:2004adc, PHOBOS:2004zne, ALICE:2022wpn}. Owing to its strongly-interacting nature, the QGP exhibits a collective response, that is effectively described by hydrodynamics, providing direct insight into its transport properties~\cite{Song:2010mg}. In non-central collisions of identical nuclei, the average overlap region is asymmetric and approximately elliptical in the transverse plane, while event-by-event fluctuations provide an additional source of initial-state anisotropy that becomes particularly important in central collisions. Interactions among the produced partons -- and, at a later stage, among the produced hadrons -- generate anisotropic pressure gradients and establish a collective medium response, thereby converting the initial spatial asymmetry into an anisotropy in the momentum-space distribution of the emitted particles~\cite{Ollitrault:1992bk, Voloshin:1994mz, Poskanzer:1998yz}. These anisotropies are quantified by the Fourier coefficients $v_n$, which generally depend on transverse momentum \pt and the pseudorapidity $\eta$ ($\eta = -\ln[\tan(\theta/2)]$) of the produced particles, where $\theta$ is the polar angle measured with respect to the beam axis. They are obtained from the decomposition of the particle azimuthal-angle distribution with respect to the corresponding symmetry plane,
\begin{equation}
    \frac{dN}{d\varphi} \propto 1 + \sum_{n} 2v_{n}(\pt, \eta)\cos\!\left[n(\varphi-\Psi_{n})\right]\;,
\end{equation}
where $\varphi$ and \pt denote the azimuthal angle and the momentum component of the particle transverse to the beam axis, respectively, and $\Psi_{n}$ is the symmetry-plane angle of the $n$-th harmonic~\cite{ATLAS:2014ndd, ALICE:2023wdn, ALICE:2024fus}.
The second-order Fourier coefficient, \vtwo, known as elliptic flow, quantifies the elliptic component of the anisotropic emission pattern and, in non-central collisions, reflects the collective response of the medium to the initial elliptic geometry of the collision~\cite{STAR:2000ekf, STAR:2003wqp, PHENIX:2003qra, STAR:2004jwm, STAR:2008ftz, STAR:2013ayu, STAR:2015rxv, STAR:2005npq, ALICE:2010suc, ALICE:2011ab, ALICE:2016tlx, ALICE:2018yph, ALICE:2014wao, ALICE:2016cti, ALICE:2016ccg, ATLAS:2012at, CMS:2013wjq}.
The associated anisotropy extends over a wide range in pseudorapidity ($\eta$), giving rise to the characteristic long-range structure in two-particle correlations commonly referred to as the ridge~\cite{STAR:2009ngv}.
Precise measurements of $v_{n}$ of produced particles in heavy-ion collisions, together with comparisons with relativistic viscous hydrodynamic calculations, have established that the QGP is a strongly-interacting, nearly perfect fluid with a viscosity to entropy density ratio approaching the conjectured quantum lower bound of $1/4\pi$~\cite{Kovtun:2004de, Heinz:2013th, Braun-Munzinger:2015hba}.

Measurements of anisotropic flow coefficients of identified hadrons provide insight into the mechanisms governing the development of collective flow and its imprint on the hadronic final state~\cite{PHENIX:2003qra, STAR:2005npq, STAR:2008ftz, STAR:2013ayu, ALICE:2018yph, ALICE:2021ibz}. In particular, the \pt-differential \vtwo of identified particles provides sensitivity to the interplay of collective dynamics developed during the collision evolution and the transfer of partonic collectivity to final-state hadrons through hadronization. At low \pt, \vtwo\ exhibits a characteristic mass ordering, with heavier hadrons showing smaller values at a given \pt, consistent with the combined effects of radial and anisotropic collective expansion developed throughout the system evolution~\cite{Huovinen:2001cy}. At intermediate \pt, a clear separation between baryons and mesons appears, commonly referred to as baryon--meson grouping and splitting~\cite{PHENIX:2006dpn, STAR:2013ayu, STAR:2015rxv, ALICE:2024vzv}. Within the quark-recombination scenario, this grouping and splitting arises from hadron formation via the recombination of constituent quarks and is regarded as a consequence of hadronization from a collectively flowing partonic medium~\cite{Molnar:2003ff}.
Although hydrodynamic calculations supplemented with fragmentation can generate some baryon–meson separation through the interplay between radial flow and the transition to fragmentation, they underestimate the measured \vtwo magnitude and do not fully reproduce the observed meson–baryon grouping at intermediate \pt~\cite{Zhao:2021vmu,ALICE:2022zks}.

Several features traditionally associated with collective behavior in heavy-ion collisions have also been observed in small collision systems at the LHC. These include the emergence of long-range multi-particle azimuthal correlations~\cite{CMS:2010ifv, ALICE:2012eyl, ATLAS:2012cix, ALICE:2013snk, ATLAS:2015hzw, Grosse-Oetringhaus:2024bwr}; an enhancement of strange-hadron production with increasing multiplicity~\cite{ALICE:2016fzo, ALICE:2019avo}; an enhancement of baryon-to-meson ratios~\cite{ALICE:2022wpn}, and $v_{2}$ grouping and splitting at intermediate $p_{\rm T}$, as reported in pp and p–Pb collision at high multiplicity~\cite{CMS:2018loe, ALICE:2019zfl, ALICE:2024vzv, CMS:2016fnw}. Such observations have stimulated extensive theoretical efforts, with interpretations ranging from initial-state momentum correlations to final-state collective dynamics described by hydrodynamic expansion and quark recombination (also referred to as quark coalescence)~\cite{Bozek:2011if, He:2015hfa, Zhou:2015iba, Weller:2017tsr, Zhao:2017rgg, Bierlich:2017vhg, Kurkela:2018ygx, Schenke:2020mbo, Schenke:2022mjv}. However, in pp and p--Pb collisions, the initial conditions are not well constrained because of uncertainties in the proton’s spatial and subnucleonic structure, together with event-by-event fluctuations of these internal configurations~\cite{Mantysaari:2017cni, Mantysaari:2022ffw}. Consequently, the underlying origin of the observed flow-like signatures remains an open question.

In this context, collisions of light ions provide a unique opportunity to investigate collective phenomena in a small nuclear system with a well-constrained initial geometry~\cite{Lim:2018huo, Brewer:2021kiv, Giacalone:2024luz}. In terms of final-state multiplicity, OO collisions bridge the high-multiplicity proton-induced and heavy-ion regimes while remaining a small nuclear system in terms of the size of the colliding nuclei. Measurements of unidentified charged-particle flow in OO collisions have already revealed evidence of collective behavior consistent with hydrodynamic model calculations~\cite{ALICE:2025luc, ATLAS:2025nnt, CMS:2025tga}. However, a complete understanding of the origin of the observed collectivity and how the initial-state anisotropies are translated into the hadronic final state during hadronization requires measurements of the flow of identified hadrons. The $\phi$ meson is particularly well suited for this purpose because its mass (\unit[1.020]{\GeVmass}) is comparable to that of the $\Lambda$ baryon (\unit[1.115]{\GeVmass}), while it has mesonic quantum numbers~\cite{ParticleDataGroup:2024cfk}. Owing to its relatively long lifetime (\unit[$\sim 46$]{fm/$c$}), which substantially exceeds the few-fm/$c$ timescale characteristic of the hadronic phase, the $\phi$ meson decays predominantly outside the hadronic medium and is therefore expected to be only weakly affected by rescattering of its decay products in this latter phase or by regeneration through hadronic interactions. Together, its mass comparable to that of the $\Lambda$ baryon, mesonic quantum numbers, and reduced sensitivity to late hadronic interactions make the $\phi$-meson \vtwo\ well suited for disentangling mass-dependent effects associated with collective expansion from particle-type-dependent effects arising during hadronization~\cite{Shor:1984ui}. Measurements of the $\phi$-meson \vtwo, therefore, provide a critical and previously missing probe of collectivity in small collision systems~\cite{ALICE:2024vzv}. When compared with corresponding measurements of baryons and lighter mesons, such as $\Lambda$($\overline{\Lambda}$), and \kzero, they provide a stringent test of hadronization via quark recombination and a robust benchmark for partonic collectivity~\cite{PHENIX:2006dpn, Fries:2008hs, ALICE:2022wpn}. These measurements enable a data-driven assessment of whether the characteristic signatures of partonic collectivity persist in light-ion collisions.

This Letter reports the first measurement of the \pt-differential \vtwo\ of \KS, $\phi$, and $\Lambda$($\overline{\Lambda}$) hadrons in OO collisions at \snn = 5.36~TeV. The main results are presented in two centrality classes defined using the charged-particle multiplicity at midrapidity, \ntrk, while complementary results obtained with a forward-rapidity multiplicity estimator are provided in the Supplemental Material. Results labeled as a given particle species correspond to the combined measurement of particles and antiparticles, when applicable.

The ALICE apparatus, as was operated during the LHC Run 3 period, is described in detail in Ref.~\cite{ALICE:2023udb}. Event selection makes use of the Fast Interaction Trigger (FIT) system~\cite{Slupecki:2022fch}, which includes the FT0A ($3.5<\eta<4.9$) and FT0C ($-3.3<\eta<-2.1$) arrays. The FT0 timing information is used in the event-quality selection, while the FT0C signal provides the forward-rapidity multiplicity estimator employed in this analysis. The reconstructed primary-vertex (PV) position along the beam direction is required to lie within \unit[7]{cm} from the nominal interaction point. Pileup events are rejected by removing bunch crossings with more than one reconstructed vertex. Additional event-quality selections reject collisions with an inconsistency between the longitudinal vertex positions reconstructed from central-barrel tracks and from the FT0 timing. In total, approximately \unit[3]{billion} OO collisions are selected for analysis. Selected events are further classified into centrality classes of 0--10\%, 10--40\%, and 65--100\%, which provide a proxy for the collision geometry. The classes are determined either from the charged-particle multiplicity at midrapidity, \ntrk, measured using tracks with $0.5 < \pt < 8.0$~\GeVc\ and $|\eta|<0.8$, or from the FT0C signal at forward rapidity. The analysis is performed in the 0--10\% and 10--40\% centrality classes, with the 0--10\% class corresponding to the highest-multiplicity events. Charged-particle tracks at midrapidity used for centrality estimation as well as the correlation analysis are reconstructed using information from the Inner Tracking System (ITS)~\cite{ALICE:2013nwm}, the Time Projection Chamber (TPC)~\cite{ALICETPC:2020ann}, and the Time-Of-Flight (TOF) detector~\cite{ALICE:2023udb, ALICE:2025tba} within the pseudorapidity range $|\eta|<0.8$. High-quality tracks are selected by requiring at least 50 associated TPC clusters, more than 80 crossed rows in the TPC, and at least five associated ITS clusters. In addition, a $\chi^{2}$ per degree of freedom smaller than 2.5 is required for the track fit to the TPC space points. Charged-particle identification for the resonance reconstruction discussed below is performed using the specific energy loss measured in the TPC and, when available, the time-of-flight information from the TOF detector. Tracks are required to have a specific energy loss within four standard deviations of the expected mean value for a given particle species over the full momentum range. When TOF information is available, an additional selection based on the deviation of the measured time-of-flight from the expected value is applied. This TOF selection is progressively tightened with increasing momentum, and tracks are required to satisfy $\beta$ $>$ 0.5 to suppress mismatched TOF associations.

The $\phi$ meson is reconstructed via the invariant mass of its decay products in the hadronic decay channel $\phi \rightarrow \mathrm{K}^{+}\mathrm{K}^{-}$~\cite{ParticleDataGroup:2024cfk}. Contamination from secondary particles arising from weak decays and interactions with the detector material is reduced by requiring the distance of closest approach (DCA) of the track to the PV to be smaller than \unit[0.1]{cm} in both the longitudinal and transverse directions. The \kzero\ and $\Lambda$ candidates are reconstructed in their weak decay channels, $\kzero \rightarrow \pi^{+}\pi^{-}$ and $\Lambda \rightarrow \mathrm{p}\pi^{-}$ ($\overline{\Lambda} \rightarrow \overline{\mathrm{p}}\pi^{+}$)~\cite{ ParticleDataGroup:2024cfk}. The elliptic-flow coefficient \vtwo\ of \kzero, $\phi$, and $\Lambda$ is measured in the transverse-momentum interval $0.5 < \pt < 8.0$~GeV/$c$. The candidate daughter tracks are required to satisfy $|\eta| < 0.8$, $\pt > 0.2$~GeV/$c$ and the track-quality criteria described above. For \kzero\ and $\Lambda$ candidates, additional selections on the distinctive V-shaped decay topology are applied in order to suppress combinatorial background~\cite{ALICE:2013cdo}. These include a requirement on the cosine of the angle between the flight-line and the reconstructed momentum to be greater than 0.993, a decay radius between 1.2 and \unit[30]{cm}, and a proper decay length requirement of $c\tau <$ \unit[20]{cm} for \kzero\ and $c\tau <$ \unit[30]{cm} for $\Lambda$. For \kzero\ candidates, a selection based on Armenteros–Podolanski variables as described in Refs.~\cite{ALICE:2013cdo, Podolanski:1954dyc} is also applied. Further background suppression is achieved by imposing requirements on the DCA to PV, including a minimum pion DCA of \unit[0.1]{cm} for \kzero\ candidates, and minimum proton and pion DCAs of \unit[0.07]{cm} and \unit[0.2]{cm}, respectively, for $\Lambda$ candidates. The DCA between the daughter tracks is required to be smaller than \unit[0.8]{cm}.

The \vtwo\ of strange hadrons is measured using the two-particle azimuthal correlation technique~\cite{STAR:2009ngv, ALICE:2012eyl}, which measures the correlation between particle pairs as a function of their azimuthal-angle and pseudorapidity differences, $\Delta\varphi$ and $\Delta\eta$, respectively, after correcting for detector-acceptance effects using mixed events and for reconstruction efficiencies obtained from PYTHIA-based Monte Carlo simulations. Short-range correlations arising from non-flow effects, i.e. correlations not associated with the common symmetry plane, such as resonance decays and near-side jet fragmentation, are suppressed by requiring a pseudorapidity separation of $|\Delta\eta|>1.2$ between the particle of interest (POI) and the reference particles, defined as unidentified charged hadrons within \pt interval of $0.2-8.0$~\GeVc and $|\eta|$ < 0.8. After this selection, the projected distributions $Y(\Delta\varphi)$  are analyzed using a low-multiplicity template-fit method~\cite{ATLAS:2015hzw}. In this method $Y(\Delta\varphi)$, measured in high-multiplicity events, is described by two components. The first is a scaled reference distribution from low-multiplicity events, which is dominated by residual non-flow contributions. The second is a Fourier series in $\Delta\varphi$ that accounts for the flow harmonics. The low-multiplicity reference corresponds to the 65--100\% \ntrk class. The fit yields the two-particle azimuthal correlation coefficient $V_{2\Delta}^{\mathrm{POI}\text{--}h}(p_{\mathrm{T}})$, which describes the correlations between the POI and unidentified charged hadrons. The elliptic-flow coefficient of the POI is then obtained by assuming factorization of the residual two-particle correlation, after suppression of non-flow contributions through the $|\Delta\eta|>1.2$ requirement and the low-multiplicity template fit, into single-particle flow coefficients and is calculated as
\begin{equation}
    v_{2}^{\mathrm{POI}}(p_{\mathrm{T}})=
    \frac{V_{2\Delta}^{\mathrm{POI}\text{--}h}(p_{\mathrm{T}})}
    {\sqrt{V_{2\Delta}^{h\text{--}h}}}\,,
\end{equation}
where $V_{2\Delta}^{h\text{--}h}$ denotes the two-particle azimuthal correlation coefficient measured between two unidentified charged hadrons in the same centrality class.

For hadrons reconstructed from their decay products (\KS, $\Lambda$, and $\phi$), the coefficient $V_{2\Delta}^{\mathrm{POI}\text{--}h}$ is extracted as a function of the invariant mass $m$, and a simultaneous fit is performed to the invariant-mass distribution and to the corresponding $V_{2\Delta}^{\mathrm{POI}\text{--}h}(m)$. The invariant-mass dependence of the two-particle azimuthal correlation coefficient is described as,
\begin{equation}
    V_{2\Delta}^{\mathrm{POI}\text{--}h}(m)=
    V_{2\Delta}^{\mathrm{sig}}
    \frac{N_{\mathrm{sig}}(m)}{N_{\mathrm{sig}}(m)+N_{\mathrm{bkg}}(m)}
    +
    V_{2\Delta}^{\mathrm{bkg}}(m)
    \frac{N_{\mathrm{bkg}}(m)}{N_{\mathrm{sig}}(m)+N_{\mathrm{bkg}}(m)},
\end{equation}
where $N_{\mathrm{sig}}(m)$ and $N_{\mathrm{bkg}}(m)$ are the invariant-mass dependent signal and background yields obtained within the same fit. The parameter $V_{2\Delta}^{\mathrm{sig}}$ denotes the two-particle elliptic-flow coefficient of the signal. The background contribution $V_{2\Delta}^{\mathrm{bkg}}(m)$ is parameterized with a second-order polynomial to describe the smooth mass dependence of the combinatorial background.
The invariant-mass signal component is modeled with a Voigtian function combined with a threshold background for the $\phi$ meson~\cite{ALICE:2018yph}, while for the \KS\ and $\Lambda$ hadrons it is described by a Gaussian signal peak and a third-order polynomial background~\cite{ALICE:2025cqy}. The signal coefficient $V_{2\Delta}^{\mathrm{sig}}$ extracted from the simultaneous fit is then used to determine the elliptic-flow coefficient $v_{2}^{\mathrm{POI}}(p_{\mathrm{T}})$.

\begin{figure}[H]
    \centering
    \includegraphics[width=0.6\textwidth]{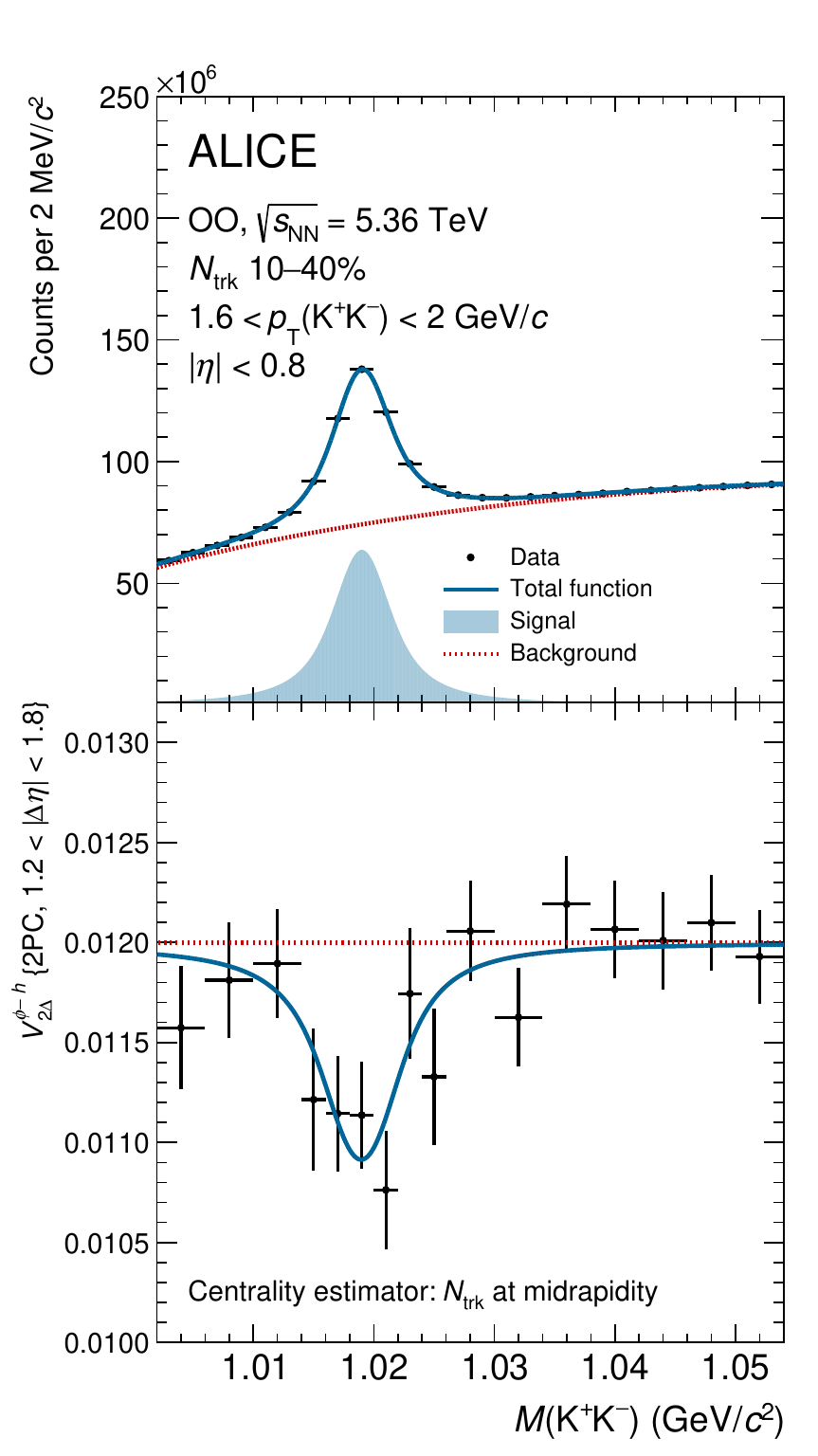}
    \caption{Simultaneous fit of the invariant-mass distribution (top panel) and the elliptic-flow coefficient, $V_{2\Delta}^{\phi\text{-}h}$ (bottom panel), as a function of the invariant mass of oppositely-charged kaon pairs in the transverse-momentum interval $1.6 < \pt < 2$~GeV/$c$ in OO collisions at \snn = 5.36~TeV for the 10--40\% \ntrk class. The total fit function is shown as a solid blue line, while the combinatorial background contribution to the invariant-mass distribution is indicated by the red dashed line.}
    \label{fig:fig1}
\end{figure}

An example of the invariant-mass dependent extraction procedure is shown in Fig.~\ref{fig:fig1}, which presents the invariant-mass distribution of oppositely-charged kaon pairs together with the corresponding two-particle azimuthal correlation coefficient, $V_{2\Delta}^{\phi\text{-}h}$, as a function of invariant mass in the transverse-momentum interval $1.6 < \pt < 2$~GeV/$c$. The results are presented for the 10--40\% centrality class. A distinct $\phi$-meson peak is observed near the nominal mass, accompanied by a depletion of $V_{2\Delta}^{\phi\text{-}h}$ relative to the background-dominated regions due to the smaller signal correlation coefficient.

Systematic uncertainties on the measured \vtwo\ are evaluated by varying the analysis conditions related to event selection, track reconstruction, particle identification, signal extraction, and the treatment of residual non-flow correlations. Uncertainties related to event selection are assessed by varying the accepted PV position. Track-reconstruction uncertainties are evaluated by modifying the DCA requirements, the number of TPC crossed rows, and the track-fit quality criteria. Particle-identification uncertainties are estimated by varying the particle-identification selections on the daughter tracks of \KS, $\Lambda$, and $\phi$ candidates. Uncertainties associated with signal extraction for \KS, $\Lambda$, and $\phi$ hadrons are evaluated by varying the invariant-mass fit ranges and the functional forms used to describe the signal and background components. For \KS\ and $\Lambda$ candidates, additional contributions are obtained by varying the topological selection criteria.
The uncertainty associated with the treatment of residual non-flow correlations is estimated by varying the pseudorapidity separation and the definition of the low-multiplicity reference used in the template-fit procedure. All combinations of these variations are considered within one source, and the corresponding systematic uncertainty is quantified from the root-mean-square of the resulting spread of \vtwo\ values with respect to the default configuration. The stability of the measurement with respect to the non-flow subtraction method is further cross-checked using the low-multiplicity-subtraction technique~\cite{ALICE:2012eyl} and Monte Carlo closure tests of the template-fit method. The individual contributions are then combined to obtain the total systematic uncertainty by summing uncorrelated sources in quadrature, while correlated contributions—such as those arising from the PV selection and the treatment of residual non-flow correlations—are treated coherently. The total systematic uncertainty is \pt-dependent and largely independent of centrality. The magnitude of the individual systematic sources are provided in the Supplementary Material~\ref{sec:SM1}.

Figure~\ref{fig:fig2} shows the \pt\ dependence of \vtwo for \kzero, $\Lambda$, and $\phi$ hadrons for the two centrality classes, 0--10\% and 10--40\%. This constitutes the first measurement of the \vtwo\ of the $\phi$ meson in a small collision system. At low transverse momentum, a clear mass ordering of \vtwo\ is observed, with $v_{2}(\kzero) > v_{2}(\Lambda) \approx v_{2}(\phi)$. The combined significance of this ordering is computed for $\pt <$~\unit[2.2]{\GeVc} taking into account the correlations of the uncertainties and expressed in the number of $\sigma$.
In both centrality classes, the separation between \kzero\ and $\Lambda$ exceeds $10\sigma$, while the difference between \kzero\ and $\phi$ exceeds $5\sigma$, demonstrating that the observed mass ordering is significant. Such a hierarchy is consistent with expectations from a common expansion velocity field, as in hydrodynamic calculations, where the interplay between radial and anisotropic flow leads to a suppression of \vtwo\ for heavier hadrons at low \pt ~\cite{Huovinen:2001cy}. The observation of mass ordering among all the presented hadrons in OO collisions provides evidence for collective expansion in a small nuclear system.

\begin{figure}[H]
    \centering
    \includegraphics[width=0.9\textwidth]{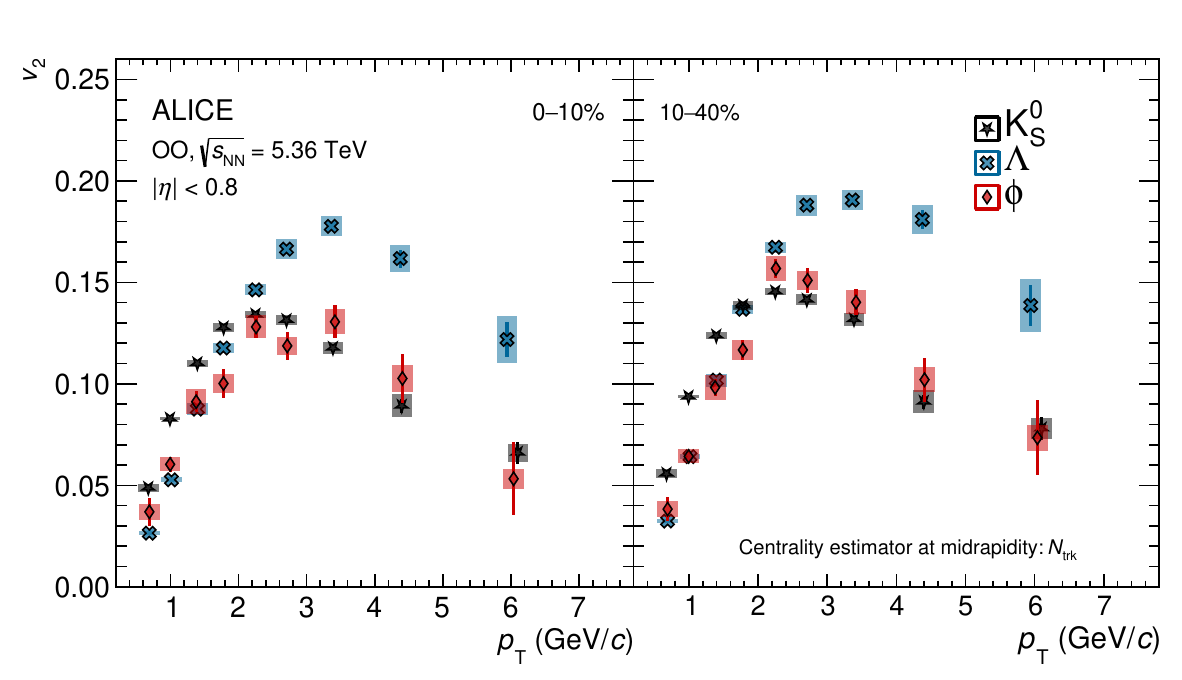}
    \caption{The results of \vtwo\ as a function of \pt for \kzero, $\Lambda$, and $\phi$ hadrons measured in OO collisions at \mbox{\snn = 5.36~TeV} for two centrality classes. Statistical and systematic uncertainties are shown as vertical bars and shaded boxes, respectively. The data points are drawn at the \meanpt of the corresponding bin.}
    \label{fig:fig2}
\end{figure}

At intermediate transverse momentum ($\pt \gtrsim$\ \unit[3]{\GeVc}), the ordering pattern changes and a clear baryon--meson separation emerges. The $\Lambda$ baryon \vtwo exceeds that of mesons, while the $\phi$ and \kzero\ \vtwo values are compatible within uncertainties and are significantly smaller than that of the $\Lambda$. In both centrality classes, the baryon–meson splitting of \vtwo above \unit[3]{\GeVc} is quantified by separations exceeding $10\sigma$ between $\Lambda$ and $\kzero$, and more than $5\sigma$ between $\Lambda$ and $\phi$.

Such a baryon--meson splitting and grouping is a characteristic feature previously observed in heavy nucleus--nucleus collisions. It is commonly attributed to hadron formation via quark recombination in a color-deconfined medium, in which hadrons inherit the collective flow of their constituent quarks. Hydrodynamic calculations supplemented with fragmentation can generate species-dependent separation through radial-flow effects and the transition from soft to hard production, with heavier hadrons retaining a larger hydrodynamic contribution to higher \pt. However, these approaches do not fully reproduce the observed magnitude and particle-species dependence of \vtwo~\cite{Zhao:2021vmu, ALICE:2022zks, YuanyuanWang:2023vor}, while transport studies indicate no significant modification from hadronic rescattering at intermediate \pt~\cite{Tang:2023wcd}.

In this context, the observation that the $\phi$ meson exhibits \vtwo values comparable to those of the substantially lighter $\kzero$ meson, rather than to those of the similarly massive $\Lambda$ baryon, provides a model-independent qualitative indication that the intermediate-\pt ordering of identified-hadron \vtwo is not governed by hadron mass alone. Instead, the observed grouping according to particle type is qualitatively consistent with hadronization through quark recombination. The measured hadrons also exhibit an approximate common trend when both \vtwo and \pt are scaled by the number of constituent quarks, with deviations from exact scaling comparable to those previously observed in Pb--Pb collisions~\cite{ALICE:2022zks}.\\
The same qualitative features and ordering patterns are obtained using the forward-rapidity FT0C centrality estimator, indicating no significant dependence on the rapidity range of the centrality estimator.
The corresponding results are presented in the Supplementary Material~\ref{sec:SM2}.

In summary, the first measurement of elliptic flow for \kzero, $\Lambda$, and $\phi$ hadrons in OO collisions at\linebreak\mbox{\snn = 5.36~TeV} is reported. The results are presented for central and semi-central collisions, using charged-particle multiplicity estimators at midrapidity and forward rapidity. At low \pt, a pronounced mass ordering of \vtwo\ is observed, consistent with collective expansion of an evolving medium. At intermediate \pt, a significant baryon--meson splitting emerges, with the $\Lambda$ baryon exhibiting larger \vtwo\ values than mesons, while the $\phi$ meson follows the same  trend as the $\kzero$ despite its $\Lambda$-like mass. This grouping of meson flow constitutes a substantial result in small collision systems that disfavors a purely mass-driven interpretation of the observed patterns and instead supports hadronization via quark recombination from a collectively flowing partonic medium. These features are robust across centrality estimators and resemble those established in large collision systems. Together, these measurements demonstrate that the key hallmarks of partonic collectivity persist in OO collisions, completing the identified-hadron flow picture in small nuclear systems and extending evidence for QGP-like behavior to the smallest nucleus--nucleus collisions accessible at the LHC.


\newenvironment{acknowledgement}{\relax}{\relax}
\begin{acknowledgement}
    \section*{Acknowledgements}
    \input{fa_2026-09-07_Opt_C.tex}
\end{acknowledgement}

\bibliographystyle{utphys}   
\bibliography{bibliography}

\newpage
\appendix

\section{Supplementary material} \label{sec:SM}

\subsection{Systematic uncertainties} \label{sec:SM1}

Table~\ref{tab:relsys} summarizes the relative systematic uncertainties on the measured \vtwo\ of the analyzed hadrons in the two centrality classes at midrapidity, 0--10\% and 10--40\%. For all systematic sources, no significant variation as a function of \pt is observed; therefore, \pt-averaged values are quoted. The only exception is the non-flow contribution. The corresponding uncertainty is estimated by varying the pseudorapidity gap between the particle of interest and reference hadrons in the range $1 < |\Delta\eta| < 1.4$, and varying the low-multiplicity class used for the template estimation from 65--100\% to 55--100\%. Different combinations of $\Delta\eta$ selections and low-multiplicity templates are used to evaluate the non-flow subtraction uncertainty. The relative systematic uncertainty associated with non-flow subtraction increases with \pt; therefore, a \pt-dependent systematic uncertainty is used. Among the considered sources, the primary vertex selection and the non-flow subtraction introduce common correlated uncertainties across particle species. The total correlated systematic uncertainty common to all particle species is estimated to range from 0.5\% to 5.7\%.

\begin{table}[htb]
    \centering
    \caption{Relative systematic uncertainties of \vtwo for \KS, $\Lambda(\overline{\Lambda})$, and $\phi$ hadrons in \OO collisions at \snn~=~\unit[5.36]{TeV} for the centrality classes, 0--10\% and 10--40\%. The quoted ranges indicate the minimum and maximum variation as a function of \pt for the two centrality classes.}
    \label{tab:relsys}
    \begin{tabular}{l|ccc}
        \toprule
        Sources of systematic uncertainty & \textbf{$\mathrm{K^0_S}$} & \textbf{$\Lambda(\overline{\Lambda})$} & \textbf{$\phi$} \\
        \midrule
Primary vertex & 0.3--0.6\% & 0.6--0.8\% & 0.6--1\% \\
Track selections & negl. & negl. & 0.6--0.8\% \\
Topological selections & 0.5--0.6\% & 0.8--1.2\% & -- \\
Particle identification & 0.2--0.3\% & 0.3--0.4\% & 0.5--0.6\% \\
Signal extraction & 0.07--0.18\% & 0.2--0.4\% & 2--3.2\% \\
Non-flow subtraction method & 0.5--6.3\% & 0.5--8.8\% & 2--9\% \\
        \bottomrule
    \end{tabular}
\end{table}

\newpage
\subsection{Results with the forward-rapidity centrality estimator} \label{sec:SM2}

\begin{figure}[htb]
    \centering
    \includegraphics[width=0.9\textwidth]{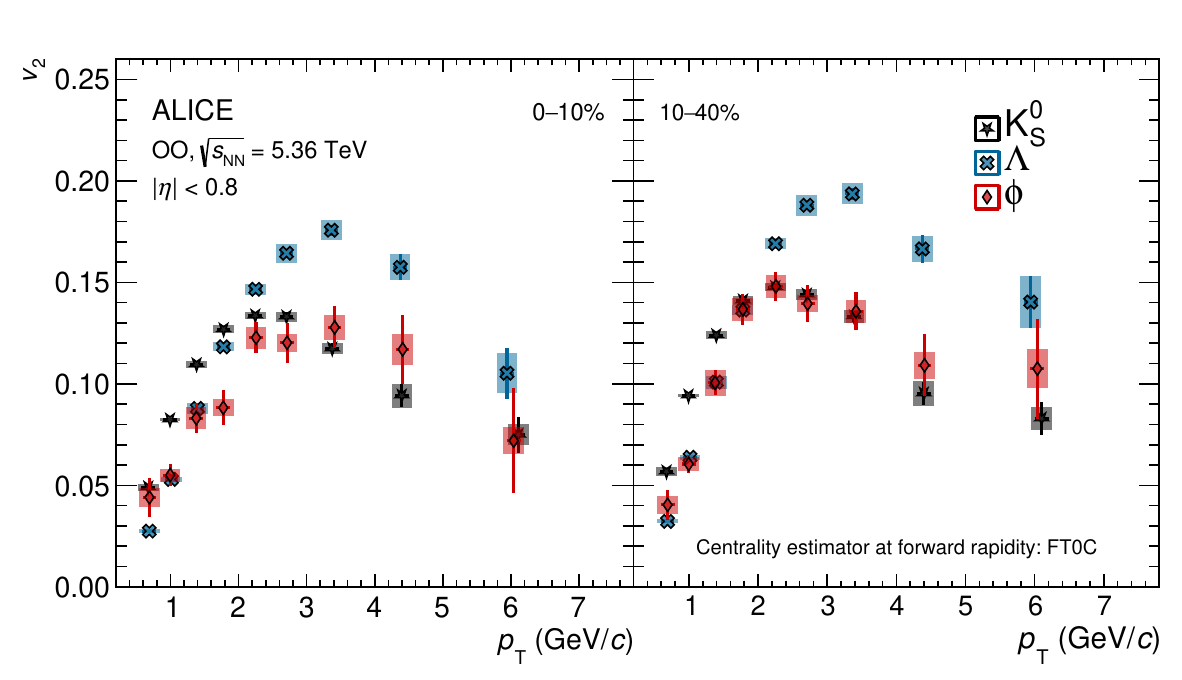}
    \caption{The value of \vtwo\ as a function of \pt\ for \kzero, $\Lambda$, and $\phi$ hadrons measured in \OO collisions at $\sqrt{s_{\mathrm{NN}}}=5.36$~TeV for the 0--10\% (left panel) and 10--40\% (right panel) centrality classes, defined using the charged-particle multiplicity measured with the forward FT0C detector. Statistical and systematic uncertainties are shown as vertical bars and shaded boxes, respectively. The data points are drawn at the \meanpt of the corresponding bin.}
    \label{fig:v0FT0C}
\end{figure}

Figure~\ref{fig:v0FT0C} presents the \vtwo\ of \kzero, $\Lambda$, and $\phi$ hadrons in \OO collisions at \snn~=~\unit[5.36]{TeV} for the 0–10\% and 10–40\% centrality classes, defined using the charged-particle multiplicity measured with the forward FT0C detector~\cite{ALICE:2023udb}, covering the pseudorapidity range  $-3.3 < \eta < -2.1$. Similarly to the results with a midrapidity centrality estimator, a mass ordering is observed at low \pt, while a separation between baryons and mesons is visible at intermediate \pt.
The combined significances are computed for $\pt <$~\unit[2.2]{\GeVc} and both centrality classes. The separation between \kzero\ and $\Lambda$ exceeds $10\sigma$, while the difference between \kzero\ and $\phi$ exceeds $5\sigma$. For $\pt >$~\unit[3]{\GeVc}, in both centrality classes, the baryon–meson splitting of \vtwo above \unit[3]{\GeVc} is quantified by separations exceeding $10\sigma$ between $\Lambda$ and $\kzero$. For $\Lambda$ and $\phi$ it is about $4.5\sigma$ in 0--10\% and more than $5\sigma$ in the 10--40\% class.

\section{The ALICE Collaboration}
\label{app:collab}
\input{Alice_Authorlist_2026-09-07_Opt_C.tex}
\end{document}

%% file: commands.tex
%

\newcommand{\pp}           {pp\xspace}
\newcommand{\OO}           {OO\xspace}
\newcommand{\ppbar}        {\mbox{$\mathrm {p\overline{p}}$}\xspace}
\newcommand{\XeXe}         {\mbox{Xe--Xe}\xspace}
\newcommand{\PbPb}         {\mbox{Pb--Pb}\xspace}
\newcommand{\pA}           {\mbox{pA}\xspace}
\newcommand{\pPb}          {\mbox{p--Pb}\xspace}
\newcommand{\AuAu}         {\mbox{Au--Au}\xspace}
\newcommand{\dAu}          {\mbox{d--Au}\xspace}

\newcommand{\s}            {\ensuremath{\sqrt{s}}\xspace}
\newcommand{\snn}          {\ensuremath{\sqrt{s_{\mathrm{NN}}}}\xspace}
\newcommand{\pt}           {\ensuremath{p_{\rm T}}\xspace}
\newcommand{\meanpt}       {$\langle p_{\mathrm{T}}\rangle$\xspace}
\newcommand{\ycms}         {\ensuremath{y_{\rm CMS}}\xspace}
\newcommand{\ylab}         {\ensuremath{y_{\rm lab}}\xspace}
\newcommand{\etarange}[1]  {\mbox{$\left | \eta \right |~<~#1$}}
\newcommand{\yrange}[1]    {\mbox{$\left | y \right |~<~#1$}}
\newcommand{\dndy}         {\ensuremath{\mathrm{d}N_\mathrm{ch}/\mathrm{d}y}\xspace}
\newcommand{\dndeta}       {\ensuremath{\mathrm{d}N_\mathrm{ch}/\mathrm{d}\eta}\xspace}
\newcommand{\avdndeta}     {\ensuremath{\langle\dndeta\rangle}\xspace}
\newcommand{\dNdy}         {\ensuremath{\mathrm{d}N_\mathrm{ch}/\mathrm{d}y}\xspace}
\newcommand{\Npart}        {\ensuremath{N_\mathrm{part}}\xspace}
\newcommand{\Ncoll}        {\ensuremath{N_\mathrm{coll}}\xspace}
\newcommand{\nch}        {\ensuremath{N_\mathrm{ch}}\xspace}
\newcommand{\ntrk}        {\ensuremath{N_\mathrm{trk}}\xspace}
\newcommand{\dEdx}         {\ensuremath{\textrm{d}E/\textrm{d}x}\xspace}
\newcommand{\RpPb}         {\ensuremath{R_{\rm pPb}}\xspace}

\newcommand{\nineH}        {$\sqrt{s}~=~0.9$~Te\kern-.1emV\xspace}
\newcommand{\seven}        {$\sqrt{s}~=~7$~Te\kern-.1emV\xspace}
\newcommand{\twoH}         {$\sqrt{s}~=~0.2$~Te\kern-.1emV\xspace}
\newcommand{\twosevensix}  {$\sqrt{s}~=~2.76$~Te\kern-.1emV\xspace}
\newcommand{\five}         {$\sqrt{s}~=~5.02$~Te\kern-.1emV\xspace}
\newcommand{\twosevensixnn}{$\sqrt{s_{\mathrm{NN}}}~=~2.76$~Te\kern-.1emV\xspace}
\newcommand{\fivenn}       {$\sqrt{s_{\mathrm{NN}}}~=~5.02$~Te\kern-.1emV\xspace}
\newcommand{\LT}           {L{\'e}vy-Tsallis\xspace}
\newcommand{\GeVc}         {Ge\kern-.1emV/$c$\xspace}
\newcommand{\MeVc}         {Me\kern-.1emV/$c$\xspace}
\newcommand{\TeV}          {Te\kern-.1emV\xspace}
\newcommand{\GeV}          {Ge\kern-.1emV\xspace}
\newcommand{\MeV}          {Me\kern-.1emV\xspace}
\newcommand{\GeVmass}      {Ge\kern-.2emV/$c^2$\xspace}
\newcommand{\MeVmass}      {Me\kern-.2emV/$c^2$\xspace}
\newcommand{\lumi}         {\ensuremath{\mathcal{L}}\xspace}

\newcommand{\ITS}          {\rm{ITS}\xspace}
\newcommand{\TOF}          {\rm{TOF}\xspace}
\newcommand{\ZDC}          {\rm{ZDC}\xspace}
\newcommand{\ZDCs}         {\rm{ZDCs}\xspace}
\newcommand{\ZNA}          {\rm{ZNA}\xspace}
\newcommand{\ZNC}          {\rm{ZNC}\xspace}
\newcommand{\SPD}          {\rm{SPD}\xspace}
\newcommand{\SDD}          {\rm{SDD}\xspace}
\newcommand{\SSD}          {\rm{SSD}\xspace}
\newcommand{\TPC}          {\rm{TPC}\xspace}
\newcommand{\TRD}          {\rm{TRD}\xspace}
\newcommand{\VZERO}        {\rm{V0}\xspace}
\newcommand{\VZEROA}       {\rm{V0A}\xspace}
\newcommand{\VZEROC}       {\rm{V0C}\xspace}
\newcommand{\Vdecay} 	   {\ensuremath{V^{0}}\xspace}

\newcommand{\ee}           {\ensuremath{e^{+}e^{-}}} 
\newcommand{\pip}          {\ensuremath{\pi^{+}}\xspace}
\newcommand{\pim}          {\ensuremath{\pi^{-}}\xspace}
\newcommand{\pipm}          {\ensuremath{\pi^{\pm}}\xspace}
\newcommand{\kap}          {\ensuremath{\rm{K}^{+}}\xspace}
\newcommand{\kam}          {\ensuremath{\rm{K}^{-}}\xspace}
\newcommand{\pbar}         {\ensuremath{\rm\overline{p}}\xspace}
\newcommand{\kzero}        {\ensuremath{{\rm K}^{0}_{\rm{S}}}\xspace}
\newcommand{\lmb}          {\ensuremath{\Lambda}\xspace}
\newcommand{\almb}         {\ensuremath{\overline{\Lambda}}\xspace}
\newcommand{\Om}           {\ensuremath{\Omega^-}\xspace}
\newcommand{\Mo}           {\ensuremath{\overline{\Omega}^+}\xspace}
\newcommand{\X}            {\ensuremath{\Xi^-}\xspace}
\newcommand{\Ix}           {\ensuremath{\overline{\Xi}^+}\xspace}
\newcommand{\Xis}          {\ensuremath{\Xi^{\pm}}\xspace}
\newcommand{\Oms}          {\ensuremath{\Omega^{\pm}}\xspace}
\newcommand{\degree}       {\ensuremath{^{\rm o}}\xspace}

\newcommand{\CKS}          {K$^{*}$(892)$^{\pm}$}
\newcommand{\CKSshort}          {K$^{*\pm}$}
\newcommand{\NKSshort}          {K$^{*0}$}
\newcommand{\NKS}          {K$^{*}$(892)$^{0}$}
\newcommand{\KS}{K$\mathrm{^{0}_{S} }$}
\newcommand{\ENfive}        {$\sqrt{\it{s}_{\mathrm{NN}}} =$ 5.02 TeV\xspace}
\newcommand{\ENthirteen}    {$\sqrt{\it{s}} =$ 13 TeV\xspace}
\newcommand{\ENtwo}        {$\sqrt{\it{s}_{\mathrm{NN}}} =$ 2.76 TeV\xspace}
\newcommand{\CKSDC}     {K$^{*\pm} \rightarrow \mathrm{K^{0}_{S} \pi^{\pm}}$\xspace}
\newcommand{\CKSDClong}     {K$^{*}$(892)$^{\pm} \rightarrow \mathrm{K^{0}_{S} \pi^{\pm}}$\xspace}
\newcommand{\NKSDClong}     {K$^{*}$(892)$^{0} \rightarrow \mathrm{K^{\mp} \pi^{\pm}}$\xspace}
\newcommand{\CKSDCKL}     {K$^{*\pm} \rightarrow \mathrm{K^{0} \pi^{\pm}}$\xspace}
\newcommand{\RAA}{$R_{\mathrm{AA}}$\xspace}
\newcommand{\pion}          {$\mathrm{\pi}$}
\newcommand{\fone}          {f$_{1}$(1285)}
\newcommand{\foneshort}          {f$_{1}$}
\newcommand{\foneDC}     {f$_{1} (1285) \rightarrow \mathrm{K^{0}_{S} K^{\pm}\pi^{\mp}}$\xspace}
\newcommand{\vtwo}          {$v_{2}$\xspace}
\newcommand{\vtwodelta}          {$V_{2\Delta}$}
\newcommand{\deta}          {\ensuremath{\Delta\eta}\xspace}

%% file: fa_2026-09-07_Opt_C.tex

The ALICE Collaboration would like to thank all its engineers and technicians for their invaluable contributions to the construction of the experiment and the CERN accelerator teams for the outstanding performance of the LHC complex.
The ALICE Collaboration gratefully acknowledges the resources and support provided by all Grid centres and the Worldwide LHC Computing Grid (WLCG) collaboration.
The ALICE Collaboration acknowledges the following funding agencies for their support in building and running the ALICE detector:
A. I. Alikhanyan National Science Laboratory (Yerevan Physics Institute) Foundation (ANSL), State Committee of Science and World Federation of Scientists (WFS), Armenia;
Austrian Academy of Sciences, Austrian Science Fund (FWF): [M 2467-N36] and Nationalstiftung f\"{u}r Forschung, Technologie und Entwicklung, Austria;
Ministry of Communications and High Technologies, National Nuclear Research Center, Azerbaijan;
Rede Nacional de Física de Altas Energias (Renafae), Financiadora de Estudos e Projetos (Finep), Funda\c{c}\~{a}o de Amparo \`{a} Pesquisa do Estado de S\~{a}o Paulo (FAPESP) and The Sao Paulo Research Foundation  (FAPESP), Brazil;
Bulgarian Ministry of Education and Science, within the National Roadmap for Research Infrastructures 2020-2027 (object CERN), Bulgaria;
Ministry of Education of China (MOEC) , Ministry of Science \& Technology of China (MSTC) and National Natural Science Foundation of China (NSFC), China;
Ministry of Science and Education and Croatian Science Foundation, Croatia;
Centro de Aplicaciones Tecnol\'{o}gicas y Desarrollo Nuclear (CEADEN), Cubaenerg\'{\i}a, Cuba;
Ministry of Education, Youth and Sports of the Czech Republic, Czech Republic;
The Danish Council for Independent Research | Natural Sciences, the VILLUM FONDEN and Danish National Research Foundation (DNRF), Denmark;
Helsinki Institute of Physics (HIP), Finland;
Commissariat \`{a} l'Energie Atomique (CEA) and Institut National de Physique Nucl\'{e}aire et de Physique des Particules (IN2P3) and Centre National de la Recherche Scientifique (CNRS), France;
Bundesministerium f\"{u}r Forschung, Technologie und Raumfahrt (BMFTR) and GSI Helmholtzzentrum f\"{u}r Schwerionenforschung GmbH, Germany;
National Research, Development and Innovation Office, Hungary;
Department of Atomic Energy Government of India (DAE), Department of Science and Technology, Government of India (DST), University Grants Commission, Government of India (UGC) and Council of Scientific and Industrial Research (CSIR), India;
National Research and Innovation Agency - BRIN, Indonesia;
Istituto Nazionale di Fisica Nucleare (INFN), Italy;
Japanese Ministry of Education, Culture, Sports, Science and Technology (MEXT) and Japan Society for the Promotion of Science (JSPS) KAKENHI, Japan;
Consejo Nacional de Ciencia (CONACYT) y Tecnolog\'{i}a, through Fondo de Cooperaci\'{o}n Internacional en Ciencia y Tecnolog\'{i}a (FONCICYT) and Direcci\'{o}n General de Asuntos del Personal Academico (DGAPA), Mexico;
Nederlandse Organisatie voor Wetenschappelijk Onderzoek (NWO), Netherlands;
The Research Council of Norway, Norway;
Pontificia Universidad Cat\'{o}lica del Per\'{u}, Peru;
Ministry of Science and Higher Education, National Science Centre and WUT ID-UB, Poland;
Korea Institute of Science and Technology Information and National Research Foundation of Korea (NRF), Republic of Korea;
Ministry of Education and Scientific Research, Institute of Atomic Physics, Ministry of Research and Innovation and Institute of Atomic Physics and Universitatea Nationala de Stiinta si Tehnologie Politehnica Bucuresti, Romania;
Ministerstvo skolstva, vyskumu, vyvoja a mladeze SR, Slovakia;
National Research Foundation of South Africa, South Africa;
Swedish Research Council (VR) and Knut \& Alice Wallenberg Foundation (KAW), Sweden;
European Organization for Nuclear Research, Switzerland;
Suranaree University of Technology (SUT), National Science and Technology Development Agency (NSTDA) and National Science, Research and Innovation Fund (NSRF via PMU-B B05F650021), Thailand;
Turkish Energy, Nuclear and Mineral Research Agency (TENMAK), Turkey;
National Academy of  Sciences of Ukraine, Ukraine;
Science and Technology Facilities Council (STFC), United Kingdom;
National Science Foundation of the United States of America (NSF) and United States Department of Energy, Office of Nuclear Physics (DOE NP), United States of America.
In addition, individual groups or members have received support from:
Czech Science Foundation (grant no. 23-07499S), Czech Republic;
FORTE project, reg.\ no.\ CZ.02.01.01/00/22\_008/0004632, Czech Republic, co-funded by the European Union, Czech Republic;
European Research Council (grant no. 101220549), European Union;
Deutsche Forschungs Gemeinschaft (DFG, German Research Foundation) ``Neutrinos and Dark Matter in Astro- and Particle Physics'' (grant no. SFB 1258), Germany;
CONVECS project, CUP C97H23001700002 FESR 2021-2027 program, Italy.

%% file: Alice_Authorlist_2026-09-07_Opt_C.tex
\begin{flushleft} 
\small

D.A.H.~Abdallah\,\orcidlink{0000-0003-4768-2718}\,$^{\rm 133}$, 
M.M.~Abedin$^{\rm 4}$, 
I.J.~Abualrob\,\orcidlink{0009-0005-3519-5631}\,$^{\rm 110}$, 
S.~Acharya\,\orcidlink{0000-0002-9213-5329}\,$^{\rm 49}$, 
A.B.~Adiguzel$^{\rm 90}$, 
K.~Agarwal\,\orcidlink{0000-0001-5781-3393}\,$^{\rm II,}$$^{\rm 23}$, 
G.~Aglieri Rinella\,\orcidlink{0000-0002-9611-3696}\,$^{\rm 32}$, 
L.~Aglietta\,\orcidlink{0009-0003-0763-6802}\,$^{\rm 24}$, 
N.~Agrawal\,\orcidlink{0000-0003-0348-9836}\,$^{\rm 25}$, 
Z.~Ahammed\,\orcidlink{0000-0001-5241-7412}\,$^{\rm 131}$
% ...

\section*{Affiliation Notes}

$^{\rm I}$ Deceased\\
$^{\rm II}$ Also at: INFN Trieste, Trieste, Italy\\
$^{\rm III}$ Also at: Fondazione Bruno Kessler (FBK), Trento, Italy\\
$^{\rm IV}$ Also at: Czech Technical University in Prague, Prague, Czech Republic\\
$^{\rm V}$ Also at: Dipartimento DET del Politecnico di Torino, Turin, Italy\\
$^{\rm VI}$ Also at: University College of Dublin, Dublin, Ireland\\
$^{\rm VII}$ Also at: Department of Applied Physics, Aligarh Muslim University, Aligarh, India\\
$^{\rm VIII}$ Also at: Institute of Theoretical Physics, University of Wroclaw, Wroclaw, Poland\\
$^{\rm IX}$ Also at: Facultad de Ciencias, Universidad Nacional Aut\'{o}noma de M\'{e}xico, Mexico City, Mexico\\
$^{\rm X}$ Also at: Lovely Professional University, Punjab, India
\\

\section*{Collaboration Institutes}

$^{1}$ A.I. Alikhanyan National Science Laboratory (Yerevan Physics Institute) Foundation, Yerevan, Armenia\\
$^{2}$ AGH University of Krakow, Cracow, Poland\\
$^{3}$ Bogolyubov Institute for Theoretical Physics, National Academy of Sciences of Ukraine, Kyiv, Ukraine\\
$^{4}$ Bose Institute, Department of Physics  and Centre for Astroparticle Physics and Space Science (CAPSS), Kolkata, India\\
$^{5}$ California Polytechnic State University, San Luis Obispo, California, United States\\
$^{6}$ Central China Normal University, Wuhan, China\\
$^{7}$ Centro de Aplicaciones Tecnol\'{o}gicas y Desarrollo Nuclear (CEADEN), Havana, Cuba\\
$^{8}$ Centro de Investigaci\'{o}n y de Estudios Avanzados (CINVESTAV), Mexico City and M\'{e}rida, Mexico\\
$^{9}$ Chicago State University, Chicago, Illinois, United States\\
$^{10}$ China Nuclear Data Center, China Institute of Atomic Energy, Beijing, China\\
$^{11}$ China University of Geosciences, Wuhan, China\\
$^{12}$ Chungbuk National University, Cheongju, Republic of Korea\\
$^{13}$ Comenius University Bratislava, Faculty of Mathematics, Physics and Informatics, Bratislava, Slovak Republic\\
$^{14}$ Creighton University, Omaha, Nebraska, United States\\
$^{15}$ Department of Physics, Aligarh Muslim University, Aligarh, India\\
$^{16}$ Department of Physics, Pusan National University, Pusan, Republic of Korea\\
$^{17}$ Department of Physics, Sejong University, Seoul, Republic of Korea\\
$^{18}$ Department of Physics, University of California, Berkeley, California, United States\\
$^{19}$ Department of Physics, University of Oslo, Oslo, Norway\\
$^{20}$ Department of Physics and Technology, University of Bergen, Bergen, Norway\\
$^{21}$ Dipartimento di Fisica, Universit\`{a} di Pavia, Pavia, Italy\\
$^{22}$ Dipartimento di Fisica dell'Universit\`{a} and Sezione INFN, Cagliari, Italy\\
$^{23}$ Dipartimento di Fisica dell'Universit\`{a} and Sezione INFN, Trieste, Italy\\
$^{24}$ Dipartimento di Fisica dell'Universit\`{a} and Sezione INFN, Turin, Italy\\
$^{25}$ Dipartimento di Fisica e Astronomia dell'Universit\`{a} and Sezione INFN, Bologna, Italy\\
$^{26}$ Dipartimento di Fisica e Astronomia dell'Universit\`{a} and Sezione INFN, Catania, Italy\\
$^{27}$ Dipartimento di Fisica e Astronomia dell'Universit\`{a} and Sezione INFN, Padova, Italy\\
$^{28}$ Dipartimento di Fisica `E.R.~Caianiello' dell'Universit\`{a} and Gruppo Collegato INFN, Salerno, Italy\\
$^{29}$ Dipartimento DISAT del Politecnico and Sezione INFN, Turin, Italy\\
$^{30}$ Dipartimento di Scienze MIFT, Universit\`{a} di Messina, Messina, Italy\\
$^{31}$ Dipartimento Interateneo di Fisica `M.~Merlin' and Sezione INFN, Bari, Italy\\
$^{32}$ European Organization for Nuclear Research (CERN), Geneva, Switzerland\\
$^{33}$ Faculty of Electrical Engineering, Mechanical Engineering and Naval Architecture, University of Split, Split, Croatia\\
$^{34}$ Faculty of Nuclear Sciences and Physical Engineering, Czech Technical University in Prague, Prague, Czech Republic\\
$^{35}$ Faculty of Physics, Sofia University, Sofia, Bulgaria\\
$^{36}$ Faculty of Science, P.J.~\v{S}af\'{a}rik University, Ko\v{s}ice, Slovak Republic\\
$^{37}$ Faculty of Technology, Environmental and Social Sciences, Bergen, Norway\\
$^{38}$ Frankfurt Institute for Advanced Studies, Johann Wolfgang Goethe-Universit\"{a}t Frankfurt, Frankfurt, Germany\\
$^{39}$ Fudan University, Shanghai, China\\
$^{40}$ Gauhati University, Department of Physics, Guwahati, India\\
$^{41}$ Helmholtz-Institut f\"{u}r Strahlen- und Kernphysik, Rheinische Friedrich-Wilhelms-Universit\"{a}t Bonn, Bonn, Germany\\
$^{42}$ Helsinki Institute of Physics (HIP), Helsinki, Finland\\
$^{43}$ High Energy Physics Group,  Universidad Aut\'{o}noma de Puebla, Puebla, Mexico\\
$^{44}$ Horia Hulubei National Institute of Physics and Nuclear Engineering, Bucharest, Romania\\
$^{45}$ HUN-REN Wigner Research Centre for Physics, Budapest, Hungary\\
$^{46}$ Indian Institute of Technology Bombay (IIT), Mumbai, India\\
$^{47}$ Indian Institute of Technology Indore, Indore, India\\
$^{48}$ INFN, Laboratori Nazionali di Frascati, Frascati, Italy\\
$^{49}$ INFN, Sezione di Bari, Bari, Italy\\
$^{50}$ INFN, Sezione di Bologna, Bologna, Italy\\
$^{51}$ INFN, Sezione di Cagliari, Cagliari, Italy\\
$^{52}$ INFN, Sezione di Catania, Catania, Italy\\
$^{53}$ INFN, Sezione di Padova, Padova, Italy\\
$^{54}$ INFN, Sezione di Pavia, Pavia, Italy\\
$^{55}$ INFN, Sezione di Torino, Turin, Italy\\
$^{56}$ INFN, Sezione di Trieste, Trieste, Italy\\
$^{57}$ Inha University, Incheon, Republic of Korea\\
$^{58}$ Institute for Gravitational and Subatomic Physics (GRASP), Utrecht University/Nikhef, Utrecht, Netherlands\\
$^{59}$ Institute of Experimental Physics, Slovak Academy of Sciences, Ko\v{s}ice, Slovak Republic\\
$^{60}$ Institute of Physics of the Czech Academy of Sciences, Prague, Czech Republic\\
$^{61}$ Institute of Space Science (ISS), Bucharest, Romania\\
$^{62}$ Institut f\"{u}r Kernphysik, Johann Wolfgang Goethe-Universit\"{a}t Frankfurt, Frankfurt, Germany\\
$^{63}$ Instituto de Ciencias Nucleares, Universidad Nacional Aut\'{o}noma de M\'{e}xico, Mexico City, Mexico\\
$^{64}$ Instituto de F\'{i}sica, Universidade Federal do Rio Grande do Sul (UFRGS), Porto Alegre, Brazil\\
$^{65}$ Instituto de F\'{\i}sica, Universidad Nacional Aut\'{o}noma de M\'{e}xico, Mexico City, Mexico\\
$^{66}$ iThemba LABS, National Research Foundation, Somerset West, South Africa\\
$^{67}$ Jeonbuk National University, Jeonju, Republic of Korea\\
$^{68}$ Korea Institute of Science and Technology Information, Daejeon, Republic of Korea\\
$^{69}$ Laboratoire de Physique Subatomique et de Cosmologie, Universit\'{e} Grenoble-Alpes, CNRS-IN2P3, Grenoble, France\\
$^{70}$ Lawrence Berkeley National Laboratory, Berkeley, California, United States\\
$^{71}$ Lund University Department of Physics, Division of Particle Physics, Lund, Sweden\\
$^{72}$ Marietta Blau Institute, Vienna, Austria\\
$^{73}$ Nagasaki Institute of Applied Science, Nagasaki, Japan\\
$^{74}$ Nara Women{'}s University (NWU), Nara, Japan\\
$^{75}$ National Centre for Nuclear Research, Warsaw, Poland\\
$^{76}$ National Institute of Science Education and Research, Homi Bhabha National Institute, Jatni, India\\
$^{77}$ National Nuclear Research Center, Baku, Azerbaijan\\
$^{78}$ National Research and Innovation Agency - BRIN, Jakarta, Indonesia\\
$^{79}$ Niels Bohr Institute, University of Copenhagen, Copenhagen, Denmark\\
$^{80}$ Nikhef, National institute for subatomic physics, Amsterdam, Netherlands\\
$^{81}$ Nuclear Physics Group, STFC Daresbury Laboratory, Daresbury, United Kingdom\\
$^{82}$ Nuclear Physics Institute of the Czech Academy of Sciences, Husinec-\v{R}e\v{z}, Czech Republic\\
$^{83}$ Oak Ridge National Laboratory, Oak Ridge, Tennessee, United States\\
$^{84}$ Ohio State University, Columbus, Ohio, United States\\
$^{85}$ Physics Department, Panjab University, Chandigarh, India\\
$^{86}$ Physics Department, University of Jammu, Jammu, India\\
$^{87}$ Physics Program and International Institute for Sustainability with Knotted Chiral Meta Matter (WPI-SKCM$^{2}$), Hiroshima University, Hiroshima, Japan\\
$^{88}$ Physikalisches Institut, Eberhard-Karls-Universit\"{a}t T\"{u}bingen, T\"{u}bingen, Germany\\
$^{89}$ Physikalisches Institut, Ruprecht-Karls-Universit\"{a}t Heidelberg, Heidelberg, Germany\\
$^{90}$ Physik Department, Technische Universit\"{a}t M\"{u}nchen, Munich, Germany\\
$^{91}$ Politecnico di Bari and Sezione INFN, Bari, Italy\\
$^{92}$ Research Division and ExtreMe Matter Institute EMMI, GSI Helmholtzzentrum f\"ur Schwerionenforschung GmbH, Darmstadt, Germany\\
$^{93}$ Saga University, Saga, Japan\\
$^{94}$ Saha Institute of Nuclear Physics, Homi Bhabha National Institute, Kolkata, India\\
$^{95}$ School of Physics and Astronomy, University of Birmingham, Birmingham, United Kingdom\\
$^{96}$ Secci\'{o}n F\'{\i}sica, Departamento de Ciencias, Pontificia Universidad Cat\'{o}lica del Per\'{u}, Lima, Peru\\
$^{97}$ SUBATECH, IMT Atlantique, Nantes Universit\'{e}, CNRS-IN2P3, Nantes, France\\
$^{98}$ Sungkyunkwan University, Suwon City, Republic of Korea\\
$^{99}$ Suranaree University of Technology, Nakhon Ratchasima, Thailand\\
$^{100}$ Technical University of Ko\v{s}ice, Ko\v{s}ice, Slovak Republic\\
$^{101}$ The Henryk Niewodniczanski Institute of Nuclear Physics, Polish Academy of Sciences, Cracow, Poland\\
$^{102}$ The University of Texas at Austin, Austin, Texas, United States\\
$^{103}$ Universidad Aut\'{o}noma de Sinaloa, Culiac\'{a}n, Mexico\\
$^{104}$ Universidade de S\~{a}o Paulo (USP), S\~{a}o Paulo, Brazil\\
$^{105}$ Universidade Estadual de Campinas (UNICAMP), Campinas, Brazil\\
$^{106}$ Universidade Federal do ABC, Santo Andre, Brazil\\
$^{107}$ Universitatea Nationala de Stiinta si Tehnologie Politehnica Bucuresti, Bucharest, Romania\\
$^{108}$ University of Cape Town, Cape Town, South Africa\\
$^{109}$ University of Derby, Derby, United Kingdom\\
$^{110}$ University of Houston, Houston, Texas, United States\\
$^{111}$ University of Jyv\"{a}skyl\"{a}, Jyv\"{a}skyl\"{a}, Finland\\
$^{112}$ University of Kansas, Lawrence, Kansas, United States\\
$^{113}$ University of Liverpool, Liverpool, United Kingdom\\
$^{114}$ University of Science and Technology of China, Hefei, China\\
$^{115}$ University of Silesia in Katowice, Katowice, Poland\\
$^{116}$ University of South-Eastern Norway, Kongsberg, Norway\\
$^{117}$ University of Tennessee, Knoxville, Tennessee, United States\\
$^{118}$ University of the Witwatersrand, Johannesburg, South Africa\\
$^{119}$ University of Tokyo, Tokyo, Japan\\
$^{120}$ University of Tsukuba, Tsukuba, Japan\\
$^{121}$ University of Zagreb Faculty of Science, Department of Physics, Zagreb, Croatia\\
$^{122}$ Universit\"{a}t M\"{u}nster, Institut f\"{u}r Kernphysik, M\"{u}nster, Germany\\
$^{123}$ Universit\'{e} Clermont Auvergne, CNRS/IN2P3, LPC, Clermont-Ferrand, France\\
$^{124}$ Universit\'{e} de Lyon, CNRS/IN2P3, Institut de Physique des 2 Infinis de Lyon, Lyon, France\\
$^{125}$ Universit\'{e} de Strasbourg, CNRS, IPHC UMR 7178, F-67000 Strasbourg, France, Strasbourg, France\\
$^{126}$ Universit\'{e} Paris-Saclay, Centre d'Etudes de Saclay (CEA), IRFU, D\'{e}partment de Physique Nucl\'{e}aire (DPhN), Saclay, France\\
$^{127}$ Universit\'{e}  Paris-Saclay, CNRS/IN2P3, IJCLab, Orsay, France\\
$^{128}$ Universit\`{a} degli Studi di Foggia, Foggia, Italy\\
$^{129}$ Universit\`{a} del Piemonte Orientale, Vercelli, Italy\\
$^{130}$ Universit\`{a} di Brescia, Brescia, Italy\\
$^{131}$ Variable Energy Cyclotron Centre, Homi Bhabha National Institute, Kolkata, India\\
$^{132}$ Warsaw University of Technology, Warsaw, Poland\\
$^{133}$ Wayne State University, Detroit, Michigan, United States\\
$^{134}$ Yale University, New Haven, Connecticut, United States\\
$^{135}$ Yildiz Technical University, Istanbul, Turkey\\
$^{136}$ Yonsei University, Seoul, Republic of Korea\\
$^{137}$ Affiliated with an institute formerly covered by a cooperation agreement with CERN\\
$^{138}$ Affiliated with an international laboratory covered by a cooperation agreement with CERN.\\

\end{flushleft} 

%% file: main.bbl
\providecommand{\href}[2]{#2}\begingroup\raggedright\begin{thebibliography}{10}

\bibitem{Shuryak:1980tp}
E.~V. Shuryak, ``{Quantum Chromodynamics and the Theory of Superdense
  Matter}'', \href{https://doi.org/10.1016/0370-1573(80)90105-2}{{\em Phys.
  Rept.} {\bfseries 61} (1980) 71--158}.

\bibitem{Cleymans:1985wb}
J.~Cleymans, R.~V. Gavai, and E.~Suhonen, ``{Quarks and Gluons at High
  Temperatures and Densities}'',
  \href{https://doi.org/10.1016/0370-1573(86)90169-9}{{\em Phys. Rept.}
  {\bfseries 130} (1986) 217}.

\bibitem{Schukraft:1993fcq}
J.~Schukraft, ``{Ultra-relativistic heavy-ion collisions: Searching for the
  quark-gluon plasma}'',
  \href{https://doi.org/10.1016/0375-9474(93)90613-3}{{\em Nucl. Phys. A}
  {\bfseries 553} (1993) 31--44}.

\bibitem{Bass:1998vz}
S.~A. Bass, M.~Gyulassy, H.~Stoecker, and W.~Greiner, ``{Signatures of quark
  gluon plasma formation in high-energy heavy ion collisions: A Critical
  review}'', \href{https://doi.org/10.1088/0954-3899/25/3/013}{{\em J. Phys. G}
  {\bfseries 25} (1999) R1--R57},
  \href{https://arxiv.org/abs/hep-ph/9810281}{{\ttfamily
  arXiv:hep-ph/9810281}}.

\bibitem{Heinz:2008tv}
U.~W. Heinz, ``{The Strongly coupled quark-gluon plasma created at RHIC}'',
  \href{https://doi.org/10.1088/1751-8113/42/21/214003}{{\em J. Phys. A}
  {\bfseries 42} (2009) 214003},
  \href{https://arxiv.org/abs/0810.5529}{{\ttfamily arXiv:0810.5529
  [nucl-th]}}.

\bibitem{Busza:2018rrf}
W.~Busza, K.~Rajagopal, and W.~van~der Schee, ``{Heavy Ion Collisions: The Big
  Picture, and the Big Questions}'',
  \href{https://doi.org/10.1146/annurev-nucl-101917-020852}{{\em Ann. Rev.
  Nucl. Part. Sci.} {\bfseries 68} (2018) 339--376},
  \href{https://arxiv.org/abs/1802.04801}{{\ttfamily arXiv:1802.04801
  [hep-ph]}}.

\bibitem{STAR:2005gfr}
{\bfseries STAR} Collaboration, J.~Adams {\em et~al.}, ``{Experimental and
  theoretical challenges in the search for the quark gluon plasma: The STAR
  Collaboration's critical assessment of the evidence from RHIC collisions}'',
  \href{https://doi.org/10.1016/j.nuclphysa.2005.03.085}{{\em Nucl. Phys. A}
  {\bfseries 757} (2005) 102--183},
  \href{https://arxiv.org/abs/nucl-ex/0501009}{{\ttfamily
  arXiv:nucl-ex/0501009}}.

\bibitem{PHENIX:2004vcz}
{\bfseries PHENIX} Collaboration, K.~Adcox {\em et~al.}, ``{Formation of dense
  partonic matter in relativistic nucleus-nucleus collisions at RHIC:
  Experimental evaluation by the PHENIX collaboration}'',
  \href{https://doi.org/10.1016/j.nuclphysa.2005.03.086}{{\em Nucl. Phys. A}
  {\bfseries 757} (2005) 184--283},
  \href{https://arxiv.org/abs/nucl-ex/0410003}{{\ttfamily
  arXiv:nucl-ex/0410003}}.

\bibitem{BRAHMS:2004adc}
{\bfseries BRAHMS} Collaboration, I.~Arsene {\em et~al.}, ``{Quark gluon plasma
  and color glass condensate at RHIC? The Perspective from the BRAHMS
  experiment}'', \href{https://doi.org/10.1016/j.nuclphysa.2005.02.130}{{\em
  Nucl. Phys. A} {\bfseries 757} (2005) 1--27},
  \href{https://arxiv.org/abs/nucl-ex/0410020}{{\ttfamily
  arXiv:nucl-ex/0410020}}.

\bibitem{PHOBOS:2004zne}
{\bfseries PHOBOS} Collaboration, B.~B. Back {\em et~al.}, ``{The PHOBOS
  perspective on discoveries at RHIC}'',
  \href{https://doi.org/10.1016/j.nuclphysa.2005.03.084}{{\em Nucl. Phys. A}
  {\bfseries 757} (2005) 28--101},
  \href{https://arxiv.org/abs/nucl-ex/0410022}{{\ttfamily
  arXiv:nucl-ex/0410022}}.

\bibitem{ALICE:2022wpn}
{\bfseries ALICE} Collaboration, S.~Acharya {\em et~al.}, ``{The ALICE
  experiment: a journey through QCD}'',
  \href{https://doi.org/10.1140/epjc/s10052-024-12935-y}{{\em Eur. Phys. J. C}
  {\bfseries 84} (2024) 813},
  \href{https://arxiv.org/abs/2211.04384}{{\ttfamily arXiv:2211.04384
  [nucl-ex]}}.

\bibitem{Song:2010mg}
H.~Song, S.~A. Bass, U.~Heinz, T.~Hirano, and C.~Shen, ``{200 A GeV Au+Au
  collisions serve a nearly perfect quark-gluon liquid}'',
  \href{https://doi.org/10.1103/PhysRevLett.106.192301}{{\em Phys. Rev. Lett.}
  {\bfseries 106} (2011) 192301},
  \href{https://arxiv.org/abs/1011.2783}{{\ttfamily arXiv:1011.2783
  [nucl-th]}}. [Erratum:
  \href{https://journals.aps.org/prl/abstract/10.1103/PhysRevLett.109.139904}{Phys.Rev.Lett.
  109, 139904 (2012)}].

\bibitem{Ollitrault:1992bk}
J.-Y. Ollitrault, ``{Anisotropy as a signature of transverse collective
  flow}'', \href{https://doi.org/10.1103/PhysRevD.46.229}{{\em Phys. Rev. D}
  {\bfseries 46} (1992) 229--245}.

\bibitem{Voloshin:1994mz}
S.~Voloshin and Y.~Zhang, ``{Flow study in relativistic nuclear collisions by
  Fourier expansion of Azimuthal particle distributions}'',
  \href{https://doi.org/10.1007/s002880050141}{{\em Z. Phys. C} {\bfseries 70}
  (1996) 665--672}, \href{https://arxiv.org/abs/hep-ph/9407282}{{\ttfamily
  arXiv:hep-ph/9407282}}.

\bibitem{Poskanzer:1998yz}
A.~M. Poskanzer and S.~A. Voloshin, ``{Methods for analyzing anisotropic flow
  in relativistic nuclear collisions}'',
  \href{https://doi.org/10.1103/PhysRevC.58.1671}{{\em Phys. Rev. C} {\bfseries
  58} (1998) 1671--1678},
  \href{https://arxiv.org/abs/nucl-ex/9805001}{{\ttfamily
  arXiv:nucl-ex/9805001}}.

\bibitem{ATLAS:2014ndd}
{\bfseries ATLAS} Collaboration, G.~Aad {\em et~al.}, ``{Measurement of
  event-plane correlations in $\sqrt{s_{NN}}=2.76$ TeV lead-lead collisions
  with the ATLAS detector}'',
  \href{https://doi.org/10.1103/PhysRevC.90.024905}{{\em Phys. Rev. C}
  {\bfseries 90} (2014) 024905},
  \href{https://arxiv.org/abs/1403.0489}{{\ttfamily arXiv:1403.0489 [hep-ex]}}.

\bibitem{ALICE:2023wdn}
{\bfseries ALICE} Collaboration, S.~Acharya {\em et~al.}, ``{Symmetry plane
  correlations in Pb-Pb collisions at $\sqrt{s_{\rm NN}} = 2.76$TeV}'',
  \href{https://doi.org/10.1140/epjc/s10052-023-11658-w}{{\em Eur. Phys. J. C}
  {\bfseries 83} (2023) 576},
  \href{https://arxiv.org/abs/2302.01234}{{\ttfamily arXiv:2302.01234
  [nucl-ex]}}.

\bibitem{ALICE:2024fus}
{\bfseries ALICE} Collaboration, S.~Acharya {\em et~al.}, ``{Higher-order
  symmetry plane correlations in Pb-Pb collisions at sNN=5.02TeV}'',
  \href{https://doi.org/10.1103/zx6t-29hf}{{\em Phys. Rev. C} {\bfseries 111}
  (2025) 064913}, \href{https://arxiv.org/abs/2409.04238}{{\ttfamily
  arXiv:2409.04238 [nucl-ex]}}.

\bibitem{STAR:2000ekf}
{\bfseries STAR} Collaboration, K.~H. Ackermann {\em et~al.}, ``{Elliptic flow
  in Au+Au collisions at $\sqrt{s_{\rm{NN}}}$ = 130 GeV}'',
  \href{https://doi.org/10.1103/PhysRevLett.86.402}{{\em Phys. Rev. Lett.}
  {\bfseries 86} (2001) 402--407},
  \href{https://arxiv.org/abs/nucl-ex/0009011}{{\ttfamily
  arXiv:nucl-ex/0009011}}.

\bibitem{STAR:2003wqp}
{\bfseries STAR} Collaboration, J.~Adams {\em et~al.}, ``{Particle type
  dependence of azimuthal anisotropy and nuclear modification of particle
  production in Au + Au collisions at \snn = 200 GeV}'',
  \href{https://doi.org/10.1103/PhysRevLett.92.052302}{{\em Phys. Rev. Lett.}
  {\bfseries 92} (2004) 052302},
  \href{https://arxiv.org/abs/nucl-ex/0306007}{{\ttfamily
  arXiv:nucl-ex/0306007}}.

\bibitem{PHENIX:2003qra}
{\bfseries PHENIX} Collaboration, S.~S. Adler {\em et~al.}, ``{Elliptic flow of
  identified hadrons in Au+Au collisions at \snn = 200 GeV}'',
  \href{https://doi.org/10.1103/PhysRevLett.91.182301}{{\em Phys. Rev. Lett.}
  {\bfseries 91} (2003) 182301},
  \href{https://arxiv.org/abs/nucl-ex/0305013}{{\ttfamily
  arXiv:nucl-ex/0305013}}.

\bibitem{STAR:2004jwm}
{\bfseries STAR} Collaboration, J.~Adams {\em et~al.}, ``{Azimuthal anisotropy
  in Au+Au collisions at \snn = 200 GeV}'',
  \href{https://doi.org/10.1103/PhysRevC.72.014904}{{\em Phys. Rev. C}
  {\bfseries 72} (2005) 014904},
  \href{https://arxiv.org/abs/nucl-ex/0409033}{{\ttfamily
  arXiv:nucl-ex/0409033}}.

\bibitem{STAR:2008ftz}
{\bfseries STAR} Collaboration, B.~I. Abelev {\em et~al.}, ``{Centrality
  dependence of charged hadron and strange hadron elliptic flow from
  $\sqrt{s_{\rm NN}}$ = 200 GeV Au + Au collisions}'',
  \href{https://doi.org/10.1103/PhysRevC.77.054901}{{\em Phys. Rev. C}
  {\bfseries 77} (2008) 054901},
  \href{https://arxiv.org/abs/0801.3466}{{\ttfamily arXiv:0801.3466
  [nucl-ex]}}.

\bibitem{STAR:2013ayu}
{\bfseries STAR} Collaboration, L.~Adamczyk {\em et~al.}, ``{Elliptic flow of
  identified hadrons in Au+Au collisions at \snn= 7.7-62.4 GeV}'',
  \href{https://doi.org/10.1103/PhysRevC.88.014902}{{\em Phys. Rev. C}
  {\bfseries 88} (2013) 014902},
  \href{https://arxiv.org/abs/1301.2348}{{\ttfamily arXiv:1301.2348
  [nucl-ex]}}.

\bibitem{STAR:2015rxv}
{\bfseries STAR} Collaboration, L.~Adamczyk {\em et~al.}, ``{Centrality
  dependence of identified particle elliptic flow in relativistic heavy ion
  collisions at $\sqrt{s_{NN}}$ = 7.7{\textendash}62.4 GeV}'',
  \href{https://doi.org/10.1103/PhysRevC.93.014907}{{\em Phys. Rev. C}
  {\bfseries 93} (2016) 014907},
  \href{https://arxiv.org/abs/1509.08397}{{\ttfamily arXiv:1509.08397
  [nucl-ex]}}.

\bibitem{STAR:2005npq}
{\bfseries STAR} Collaboration, J.~Adams {\em et~al.}, ``{Multi-strange baryon
  elliptic flow in Au + Au collisions at \snn = 200 GeV}'',
  \href{https://doi.org/10.1103/PhysRevLett.95.122301}{{\em Phys. Rev. Lett.}
  {\bfseries 95} (2005) 122301},
  \href{https://arxiv.org/abs/nucl-ex/0504022}{{\ttfamily
  arXiv:nucl-ex/0504022}}.

\bibitem{ALICE:2010suc}
{\bfseries ALICE} Collaboration, K.~Aamodt {\em et~al.}, ``{Elliptic flow of
  charged particles in Pb--Pb collisions at 2.76 TeV}'',
  \href{https://doi.org/10.1103/PhysRevLett.105.252302}{{\em Phys. Rev. Lett.}
  {\bfseries 105} (2010) 252302},
  \href{https://arxiv.org/abs/1011.3914}{{\ttfamily arXiv:1011.3914
  [nucl-ex]}}.

\bibitem{ALICE:2011ab}
{\bfseries ALICE} Collaboration, K.~Aamodt {\em et~al.}, ``{Higher harmonic
  anisotropic flow measurements of charged particles in Pb--Pb collisions at
  $\sqrt{s_{NN}}$ = 2.76 TeV}'',
  \href{https://doi.org/10.1103/PhysRevLett.107.032301}{{\em Phys. Rev. Lett.}
  {\bfseries 107} (2011) 032301},
  \href{https://arxiv.org/abs/1105.3865}{{\ttfamily arXiv:1105.3865
  [nucl-ex]}}.

\bibitem{ALICE:2016tlx}
{\bfseries ALICE} Collaboration, J.~Adam {\em et~al.}, ``{Pseudorapidity
  dependence of the anisotropic flow of charged particles in Pb--Pb collisions
  at $\sqrt{s_{\rm NN}}$ = 2.76 TeV}'',
  \href{https://doi.org/10.1016/j.physletb.2016.07.017}{{\em Phys. Lett. B}
  {\bfseries 762} (2016) 376--388},
  \href{https://arxiv.org/abs/1605.02035}{{\ttfamily arXiv:1605.02035
  [nucl-ex]}}.

\bibitem{ALICE:2018yph}
{\bfseries ALICE} Collaboration, S.~Acharya {\em et~al.}, ``{Anisotropic flow
  of identified particles in Pb--Pb collisions at $
  {\sqrt{s}}_{\mathrm{NN}}=5.02 $ TeV}'',
  \href{https://doi.org/10.1007/JHEP09(2018)006}{{\em JHEP} {\bfseries 09}
  (2018) 006}, \href{https://arxiv.org/abs/1805.04390}{{\ttfamily
  arXiv:1805.04390 [nucl-ex]}}.

\bibitem{ALICE:2014wao}
{\bfseries ALICE} Collaboration, B.~B. Abelev {\em et~al.}, ``{Elliptic flow of
  identified hadrons in Pb--Pb collisions at $ \sqrt{s_{\mathrm{NN}}}=2.76 $
  TeV}'', \href{https://doi.org/10.1007/JHEP06(2015)190}{{\em JHEP} {\bfseries
  06} (2015) 190}, \href{https://arxiv.org/abs/1405.4632}{{\ttfamily
  arXiv:1405.4632 [nucl-ex]}}.

\bibitem{ALICE:2016cti}
{\bfseries ALICE} Collaboration, J.~Adam {\em et~al.}, ``{Higher harmonic flow
  coefficients of identified hadrons in Pb--Pb collisions at $\sqrt{s_{\rm
  NN}}$ = 2.76 TeV}'', \href{https://doi.org/10.1007/JHEP09(2016)164}{{\em
  JHEP} {\bfseries 09} (2016) 164},
  \href{https://arxiv.org/abs/1606.06057}{{\ttfamily arXiv:1606.06057
  [nucl-ex]}}.

\bibitem{ALICE:2016ccg}
{\bfseries ALICE} Collaboration, J.~Adam {\em et~al.}, ``{Anisotropic flow of
  charged particles in Pb--Pb collisions at $\sqrt{s_{\rm NN}}=5.02$ TeV}'',
  \href{https://doi.org/10.1103/PhysRevLett.116.132302}{{\em Phys. Rev. Lett.}
  {\bfseries 116} (2016) 132302},
  \href{https://arxiv.org/abs/1602.01119}{{\ttfamily arXiv:1602.01119
  [nucl-ex]}}.

\bibitem{ATLAS:2012at}
{\bfseries ATLAS} Collaboration, G.~Aad {\em et~al.}, ``{Measurement of the
  azimuthal anisotropy for charged particle production in $\sqrt{s_{NN}}=2.76$
  TeV lead-lead collisions with the ATLAS detector}'',
  \href{https://doi.org/10.1103/PhysRevC.86.014907}{{\em Phys. Rev. C}
  {\bfseries 86} (2012) 014907},
  \href{https://arxiv.org/abs/1203.3087}{{\ttfamily arXiv:1203.3087 [hep-ex]}}.

\bibitem{CMS:2013wjq}
{\bfseries CMS} Collaboration, S.~Chatrchyan {\em et~al.}, ``{Measurement of
  Higher-Order Harmonic Azimuthal Anisotropy in PbPb Collisions at
  $\sqrt{s_{NN}}$ = 2.76 TeV}'',
  \href{https://doi.org/10.1103/PhysRevC.89.044906}{{\em Phys. Rev. C}
  {\bfseries 89} (2014) 044906},
  \href{https://arxiv.org/abs/1310.8651}{{\ttfamily arXiv:1310.8651
  [nucl-ex]}}.

\bibitem{STAR:2009ngv}
{\bfseries STAR} Collaboration, B.~I. Abelev {\em et~al.}, ``{Long range
  rapidity correlations and jet production in high energy nuclear
  collisions}'', \href{https://doi.org/10.1103/PhysRevC.80.064912}{{\em Phys.
  Rev. C} {\bfseries 80} (2009) 064912},
  \href{https://arxiv.org/abs/0909.0191}{{\ttfamily arXiv:0909.0191
  [nucl-ex]}}.

\bibitem{Kovtun:2004de}
P.~Kovtun, D.~T. Son, and A.~O. Starinets, ``{Viscosity in strongly interacting
  quantum field theories from black hole physics}'',
  \href{https://doi.org/10.1103/PhysRevLett.94.111601}{{\em Phys. Rev. Lett.}
  {\bfseries 94} (2005) 111601},
  \href{https://arxiv.org/abs/hep-th/0405231}{{\ttfamily
  arXiv:hep-th/0405231}}.

\bibitem{Heinz:2013th}
U.~Heinz and R.~Snellings, ``{Collective flow and viscosity in relativistic
  heavy-ion collisions}'',
  \href{https://doi.org/10.1146/annurev-nucl-102212-170540}{{\em Ann. Rev.
  Nucl. Part. Sci.} {\bfseries 63} (2013) 123--151},
  \href{https://arxiv.org/abs/1301.2826}{{\ttfamily arXiv:1301.2826
  [nucl-th]}}.

\bibitem{Braun-Munzinger:2015hba}
P.~Braun-Munzinger, V.~Koch, T.~Sch{\"a}fer, and J.~Stachel, ``{Properties of
  hot and dense matter from relativistic heavy ion collisions}'',
  \href{https://doi.org/10.1016/j.physrep.2015.12.003}{{\em Phys. Rept.}
  {\bfseries 621} (2016) 76--126},
  \href{https://arxiv.org/abs/1510.00442}{{\ttfamily arXiv:1510.00442
  [nucl-th]}}.

\bibitem{ALICE:2021ibz}
{\bfseries ALICE} Collaboration, S.~Acharya {\em et~al.}, ``{Anisotropic flow
  of identified hadrons in Xe-Xe collisions at $ \sqrt{s_{\mathrm{NN}}} $ =
  5.44 TeV}'', \href{https://doi.org/10.1007/JHEP10(2021)152}{{\em JHEP}
  {\bfseries 10} (2021) 152},
  \href{https://arxiv.org/abs/2107.10592}{{\ttfamily arXiv:2107.10592
  [nucl-ex]}}.

\bibitem{Huovinen:2001cy}
P.~Huovinen, P.~F. Kolb, U.~W. Heinz, P.~V. Ruuskanen, and S.~A. Voloshin,
  ``{Radial and elliptic flow at RHIC: Further predictions}'',
  \href{https://doi.org/10.1016/S0370-2693(01)00219-2}{{\em Phys. Lett. B}
  {\bfseries 503} (2001) 58--64},
  \href{https://arxiv.org/abs/hep-ph/0101136}{{\ttfamily
  arXiv:hep-ph/0101136}}.

\bibitem{PHENIX:2006dpn}
{\bfseries PHENIX} Collaboration, A.~Adare {\em et~al.}, ``{Scaling properties
  of azimuthal anisotropy in Au+Au and Cu+Cu collisions at \snn = 200 GeV}'',
  \href{https://doi.org/10.1103/PhysRevLett.98.162301}{{\em Phys. Rev. Lett.}
  {\bfseries 98} (2007) 162301},
  \href{https://arxiv.org/abs/nucl-ex/0608033}{{\ttfamily
  arXiv:nucl-ex/0608033}}.

\bibitem{ALICE:2024vzv}
{\bfseries ALICE} Collaboration, S.~Acharya {\em et~al.}, ``{Observation of
  partonic flow in proton--proton and proton--nucleus collisions}'',
  \href{https://doi.org/10.1038/s41467-025-67795-1}{{\em Nature Commun.}
  {\bfseries 17} (2026) 2585},
  \href{https://arxiv.org/abs/2411.09323}{{\ttfamily arXiv:2411.09323
  [nucl-ex]}}.

\bibitem{Molnar:2003ff}
D.~Molnar and S.~A. Voloshin, ``{Elliptic flow at large transverse momenta from
  quark coalescence}'',
  \href{https://doi.org/10.1103/PhysRevLett.91.092301}{{\em Phys. Rev. Lett.}
  {\bfseries 91} (2003) 092301},
  \href{https://arxiv.org/abs/nucl-th/0302014}{{\ttfamily
  arXiv:nucl-th/0302014}}.

\bibitem{Zhao:2021vmu}
W.~Zhao, W.~Ke, W.~Chen, T.~Luo, and X.-N. Wang, ``{From Hydrodynamics to Jet
  Quenching, Coalescence, and Hadron Cascade: A Coupled Approach to Solving the
  RAA{\ensuremath{\otimes}}v2 Puzzle}'',
  \href{https://doi.org/10.1103/PhysRevLett.128.022302}{{\em Phys. Rev. Lett.}
  {\bfseries 128} (2022) 022302},
  \href{https://arxiv.org/abs/2103.14657}{{\ttfamily arXiv:2103.14657
  [hep-ph]}}.

\bibitem{ALICE:2022zks}
{\bfseries ALICE} Collaboration, S.~Acharya {\em et~al.}, ``{Anisotropic flow
  and flow fluctuations of identified hadrons in Pb{\textendash}Pb collisions
  at $ \sqrt{s_{\textrm{NN}}} $ = 5.02 TeV}'',
  \href{https://doi.org/10.1007/JHEP05(2023)243}{{\em JHEP} {\bfseries 05}
  (2023) 243}, \href{https://arxiv.org/abs/2206.04587}{{\ttfamily
  arXiv:2206.04587 [nucl-ex]}}.

\bibitem{CMS:2010ifv}
{\bfseries CMS} Collaboration, V.~Khachatryan {\em et~al.}, ``{Observation of
  Long-Range Near-Side Angular Correlations in Proton-Proton Collisions at the
  LHC}'', \href{https://doi.org/10.1007/JHEP09(2010)091}{{\em JHEP} {\bfseries
  09} (2010) 091}, \href{https://arxiv.org/abs/1009.4122}{{\ttfamily
  arXiv:1009.4122 [hep-ex]}}.

\bibitem{ALICE:2012eyl}
{\bfseries ALICE} Collaboration, B.~Abelev {\em et~al.}, ``{Long-range angular
  correlations on the near and away side in $p$-Pb collisions at
  $\sqrt{s_{NN}}=5.02$ TeV}'',
  \href{https://doi.org/10.1016/j.physletb.2013.01.012}{{\em Phys. Lett. B}
  {\bfseries 719} (2013) 29--41},
  \href{https://arxiv.org/abs/1212.2001}{{\ttfamily arXiv:1212.2001
  [nucl-ex]}}.

\bibitem{ATLAS:2012cix}
{\bfseries ATLAS} Collaboration, G.~Aad {\em et~al.}, ``{Observation of
  Associated Near-Side and Away-Side Long-Range Correlations in $\sqrt{s_{NN}}$
  = 5.02 TeV Proton-Lead Collisions with the ATLAS Detector}'',
  \href{https://doi.org/10.1103/PhysRevLett.110.182302}{{\em Phys. Rev. Lett.}
  {\bfseries 110} (2013) 182302},
  \href{https://arxiv.org/abs/1212.5198}{{\ttfamily arXiv:1212.5198 [hep-ex]}}.

\bibitem{ALICE:2013snk}
{\bfseries ALICE} Collaboration, B.~B. Abelev {\em et~al.}, ``{Long-range
  angular correlations of $\rm \pi$, K and p in p-Pb collisions at
  $\sqrt{s_{\rm NN}}$ = 5.02 TeV}'',
  \href{https://doi.org/10.1016/j.physletb.2013.08.024}{{\em Phys. Lett. B}
  {\bfseries 726} (2013) 164--177},
  \href{https://arxiv.org/abs/1307.3237}{{\ttfamily arXiv:1307.3237
  [nucl-ex]}}.

\bibitem{ATLAS:2015hzw}
{\bfseries ATLAS} Collaboration, G.~Aad {\em et~al.}, ``{Observation of
  Long-Range Elliptic Azimuthal Anisotropies in $\sqrt{s}=$13 and 2.76 TeV $pp$
  Collisions with the ATLAS Detector}'',
  \href{https://doi.org/10.1103/PhysRevLett.116.172301}{{\em Phys. Rev. Lett.}
  {\bfseries 116} (2016) 172301},
  \href{https://arxiv.org/abs/1509.04776}{{\ttfamily arXiv:1509.04776
  [hep-ex]}}.

\bibitem{Grosse-Oetringhaus:2024bwr}
J.~F. Grosse-Oetringhaus and U.~A. Wiedemann, ``{A Decade of Collectivity in
  Small Systems}'', \href{https://arxiv.org/abs/2407.07484}{{\ttfamily
  arXiv:2407.07484 [hep-ex]}}.

\bibitem{ALICE:2016fzo}
{\bfseries ALICE} Collaboration, J.~Adam {\em et~al.}, ``{Enhanced production
  of multi-strange hadrons in high-multiplicity proton-proton collisions}'',
  \href{https://doi.org/10.1038/nphys4111}{{\em Nature Phys.} {\bfseries 13}
  (2017) 535--539}, \href{https://arxiv.org/abs/1606.07424}{{\ttfamily
  arXiv:1606.07424 [nucl-ex]}}.

\bibitem{ALICE:2019avo}
{\bfseries ALICE} Collaboration, S.~Acharya {\em et~al.}, ``{Multiplicity
  dependence of (multi-)strange hadron production in proton-proton collisions
  at $\sqrt{s}$ = 13 TeV}'',
  \href{https://doi.org/10.1140/epjc/s10052-020-7673-8}{{\em Eur. Phys. J. C}
  {\bfseries 80} (2020) 167},
  \href{https://arxiv.org/abs/1908.01861}{{\ttfamily arXiv:1908.01861
  [nucl-ex]}}.

\bibitem{CMS:2018loe}
{\bfseries CMS} Collaboration, A.~M. Sirunyan {\em et~al.}, ``{Elliptic flow of
  charm and strange hadrons in high-multiplicity pPb collisions at
  $\sqrt{s_{_\mathrm{NN}}} =$ 8.16 TeV}'',
  \href{https://doi.org/10.1103/PhysRevLett.121.082301}{{\em Phys. Rev. Lett.}
  {\bfseries 121} (2018) 082301},
  \href{https://arxiv.org/abs/1804.09767}{{\ttfamily arXiv:1804.09767
  [hep-ex]}}.

\bibitem{ALICE:2019zfl}
{\bfseries ALICE} Collaboration, S.~Acharya {\em et~al.}, ``{Investigations of
  Anisotropic Flow Using Multiparticle Azimuthal Correlations in pp, p-Pb,
  Xe-Xe, and Pb--Pb Collisions at the LHC}'',
  \href{https://doi.org/10.1103/PhysRevLett.123.142301}{{\em Phys. Rev. Lett.}
  {\bfseries 123} (2019) 142301},
  \href{https://arxiv.org/abs/1903.01790}{{\ttfamily arXiv:1903.01790
  [nucl-ex]}}.

\bibitem{CMS:2016fnw}
{\bfseries CMS} Collaboration, V.~Khachatryan {\em et~al.}, ``{Evidence for
  collectivity in pp collisions at the LHC}'',
  \href{https://doi.org/10.1016/j.physletb.2016.12.009}{{\em Phys. Lett. B}
  {\bfseries 765} (2017) 193--220},
  \href{https://arxiv.org/abs/1606.06198}{{\ttfamily arXiv:1606.06198
  [nucl-ex]}}.

\bibitem{Bozek:2011if}
P.~Bozek, ``{Collective flow in p-Pb and d-Pb collisions at TeV energies}'',
  \href{https://doi.org/10.1103/PhysRevC.85.014911}{{\em Phys. Rev. C}
  {\bfseries 85} (2012) 014911},
  \href{https://arxiv.org/abs/1112.0915}{{\ttfamily arXiv:1112.0915 [hep-ph]}}.

\bibitem{He:2015hfa}
L.~He, T.~Edmonds, Z.-W. Lin, F.~Liu, D.~Molnar, and F.~Wang, ``{Anisotropic
  parton escape is the dominant source of azimuthal anisotropy in transport
  models}'', \href{https://doi.org/10.1016/j.physletb.2015.12.051}{{\em Phys.
  Lett. B} {\bfseries 753} (2016) 506--510},
  \href{https://arxiv.org/abs/1502.05572}{{\ttfamily arXiv:1502.05572
  [nucl-th]}}.

\bibitem{Zhou:2015iba}
Y.~Zhou, X.~Zhu, P.~Li, and H.~Song, ``{Investigation of possible hadronic flow
  in $\sqrt{s_{NN}} = 5.02$ TeV $p-Pb$ collisions}'',
  \href{https://doi.org/10.1103/PhysRevC.91.064908}{{\em Phys. Rev. C}
  {\bfseries 91} (2015) 064908},
  \href{https://arxiv.org/abs/1503.06986}{{\ttfamily arXiv:1503.06986
  [nucl-th]}}.

\bibitem{Weller:2017tsr}
R.~D. Weller and P.~Romatschke, ``{One fluid to rule them all: viscous
  hydrodynamic description of event-by-event central p+p, p+Pb and Pb+Pb
  collisions at $\sqrt{s}=5.02$ TeV}'',
  \href{https://doi.org/10.1016/j.physletb.2017.09.077}{{\em Phys. Lett. B}
  {\bfseries 774} (2017) 351--356},
  \href{https://arxiv.org/abs/1701.07145}{{\ttfamily arXiv:1701.07145
  [nucl-th]}}.

\bibitem{Zhao:2017rgg}
W.~Zhao, Y.~Zhou, H.~Xu, W.~Deng, and H.~Song, ``{Hydrodynamic collectivity in
  proton{\textendash}proton collisions at 13 TeV}'',
  \href{https://doi.org/10.1016/j.physletb.2018.03.022}{{\em Phys. Lett. B}
  {\bfseries 780} (2018) 495--500},
  \href{https://arxiv.org/abs/1801.00271}{{\ttfamily arXiv:1801.00271
  [nucl-th]}}.

\bibitem{Bierlich:2017vhg}
C.~Bierlich, G.~Gustafson, and L.~L{\"o}nnblad, ``{Collectivity without plasma
  in hadronic collisions}'',
  \href{https://doi.org/10.1016/j.physletb.2018.01.069}{{\em Phys. Lett. B}
  {\bfseries 779} (2018) 58--63},
  \href{https://arxiv.org/abs/1710.09725}{{\ttfamily arXiv:1710.09725
  [hep-ph]}}.

\bibitem{Kurkela:2018ygx}
A.~Kurkela, U.~A. Wiedemann, and B.~Wu, ``{Nearly isentropic flow at sizeable
  $\eta/s$}'', \href{https://doi.org/10.1016/j.physletb.2018.06.064}{{\em Phys.
  Lett. B} {\bfseries 783} (2018) 274--279},
  \href{https://arxiv.org/abs/1803.02072}{{\ttfamily arXiv:1803.02072
  [hep-ph]}}.

\bibitem{Schenke:2020mbo}
B.~Schenke, C.~Shen, and P.~Tribedy, ``{Running the gamut of high energy
  nuclear collisions}'',
  \href{https://doi.org/10.1103/PhysRevC.102.044905}{{\em Phys. Rev. C}
  {\bfseries 102} (2020) 044905},
  \href{https://arxiv.org/abs/2005.14682}{{\ttfamily arXiv:2005.14682
  [nucl-th]}}.

\bibitem{Schenke:2022mjv}
B.~Schenke, S.~Schlichting, and P.~Singh, ``{Rapidity dependence of initial
  state geometry and momentum correlations in p+Pb collisions}'',
  \href{https://doi.org/10.1103/PhysRevD.105.094023}{{\em Phys. Rev. D}
  {\bfseries 105} (2022) 094023},
  \href{https://arxiv.org/abs/2201.08864}{{\ttfamily arXiv:2201.08864
  [nucl-th]}}.

\bibitem{Mantysaari:2017cni}
H.~M{\"a}ntysaari, B.~Schenke, C.~Shen, and P.~Tribedy, ``{Imprints of
  fluctuating proton shapes on flow in proton-lead collisions at the LHC}'',
  \href{https://doi.org/10.1016/j.physletb.2017.07.038}{{\em Phys. Lett. B}
  {\bfseries 772} (2017) 681--686},
  \href{https://arxiv.org/abs/1705.03177}{{\ttfamily arXiv:1705.03177
  [nucl-th]}}.

\bibitem{Mantysaari:2022ffw}
H.~M{\"a}ntysaari, B.~Schenke, C.~Shen, and W.~Zhao, ``{Bayesian inference of
  the fluctuating proton shape}'',
  \href{https://doi.org/10.1016/j.physletb.2022.137348}{{\em Phys. Lett. B}
  {\bfseries 833} (2022) 137348},
  \href{https://arxiv.org/abs/2202.01998}{{\ttfamily arXiv:2202.01998
  [hep-ph]}}.

\bibitem{Lim:2018huo}
S.~H. Lim, J.~Carlson, C.~Loizides, D.~Lonardoni, J.~E. Lynn, J.~L. Nagle,
  J.~D. Orjuela~Koop, and J.~Ouellette, ``{Exploring New Small System
  Geometries in Heavy Ion Collisions}'',
  \href{https://doi.org/10.1103/PhysRevC.99.044904}{{\em Phys. Rev. C}
  {\bfseries 99} (2019) 044904},
  \href{https://arxiv.org/abs/1812.08096}{{\ttfamily arXiv:1812.08096
  [nucl-th]}}.

\bibitem{Brewer:2021kiv}
J.~Brewer, A.~Mazeliauskas, and W.~van~der Schee, ``{Opportunities of OO and
  $p$O collisions at the LHC}'', in {\em {Opportunities of OO and pO collisions
  at the LHC}}.
\newblock 3, 2021.
\newblock \href{https://arxiv.org/abs/2103.01939}{{\ttfamily arXiv:2103.01939
  [hep-ph]}}.

\bibitem{Giacalone:2024luz}
G.~Giacalone {\em et~al.}, ``{Exploiting $^{20}$Ne Isotopes for Precision
  Characterizations of Collectivity in Small Systems}'',
  \href{https://doi.org/10.1103/k8rb-jgvq}{{\em Phys. Rev. Lett.} {\bfseries
  135} (2025) 012302}, \href{https://arxiv.org/abs/2402.05995}{{\ttfamily
  arXiv:2402.05995 [nucl-th]}}.

\bibitem{ALICE:2025luc}
{\bfseries ALICE} Collaboration, I.~J. Abualrob {\em et~al.}, ``{Evidence of
  nuclear geometry-driven anisotropic flow in OO and Ne$-$Ne collisions at
  $\mathbf{\sqrt{{\textit s}_{\rm\mathbf {NN}}}}$ = 5.36 TeV}'',
  \href{https://arxiv.org/abs/2509.06428}{{\ttfamily arXiv:2509.06428
  [nucl-ex]}}.

\bibitem{ATLAS:2025nnt}
{\bfseries ATLAS} Collaboration, G.~Aad {\em et~al.}, ``{Measurement of the
  azimuthal anisotropy of charged particles in \snn=5.36~TeV~$O^{16}+O^{16}$
  and $Ne^{20}+Ne^{20}$ collisions with the ATLAS detector}'',
  \href{https://doi.org/10.1103/xqxz-8bhf}{{\em Phys. Rev. C} {\bfseries 113}
  (2026) 045205}, \href{https://arxiv.org/abs/2509.05171}{{\ttfamily
  arXiv:2509.05171 [nucl-ex]}}.

\bibitem{CMS:2025tga}
{\bfseries CMS} Collaboration, A.~Hayrapetyan {\em et~al.}, ``{Observation of
  Long-Range Collective Flow in O+O and Ne+Ne Collisions and Implications for
  Nuclear Structure Studies}'', \href{https://doi.org/10.1103/26wx-tg6f}{{\em
  Phys. Rev. Lett.} {\bfseries 137} (2026) 082302},
  \href{https://arxiv.org/abs/2510.02580}{{\ttfamily arXiv:2510.02580
  [nucl-ex]}}.

\bibitem{ParticleDataGroup:2024cfk}
{\bfseries Particle Data Group} Collaboration, S.~Navas {\em et~al.}, ``{Review
  of particle physics}'',
  \href{https://doi.org/10.1103/PhysRevD.110.030001}{{\em Phys. Rev. D}
  {\bfseries 110} (2024) 030001}.

\bibitem{Shor:1984ui}
A.~Shor, ``{$\phi$ meson production as a probe of the quark gluon plasma}'',
  \href{https://doi.org/10.1103/PhysRevLett.54.1122}{{\em Phys. Rev. Lett.}
  {\bfseries 54} (1985) 1122--1125}.

\bibitem{Fries:2008hs}
R.~J. Fries, V.~Greco, and P.~Sorensen, ``{Coalescence models for hadron
  formation from quark gluon plasma}'',
  \href{https://doi.org/10.1146/annurev.nucl.58.110707.171134}{{\em Ann. Rev.
  Nucl. Part. Sci.} {\bfseries 58} (2008) 177--205},
  \href{https://arxiv.org/abs/0807.4939}{{\ttfamily arXiv:0807.4939
  [nucl-th]}}.

\bibitem{ALICE:2023udb}
{\bfseries ALICE} Collaboration, S.~Acharya {\em et~al.}, ``{ALICE upgrades
  during the LHC Long Shutdown 2}'',
  \href{https://doi.org/10.1088/1748-0221/19/05/P05062}{{\em JINST} {\bfseries
  19} (2024) P05062}, \href{https://arxiv.org/abs/2302.01238}{{\ttfamily
  arXiv:2302.01238 [physics.ins-det]}}.

\bibitem{Slupecki:2022fch}
{\bfseries ALICE} Collaboration, M.~Slupecki, ``{Fast Interaction Trigger for
  ALICE upgrade}'', \href{https://doi.org/10.1016/j.nima.2022.167021}{{\em
  Nucl. Instrum. Meth. A} {\bfseries 1039} (2022) 167021}.

\bibitem{ALICE:2013nwm}
{\bfseries ALICE} Collaboration, B.~Abelev {\em et~al.}, ``{Technical Design
  Report for the Upgrade of the ALICE Inner Tracking System}'',
  \href{https://doi.org/10.1088/0954-3899/41/8/087002}{{\em J. Phys. G}
  {\bfseries 41} (2014) 087002}.

\bibitem{ALICETPC:2020ann}
{\bfseries ALICE TPC} Collaboration, J.~Adolfsson {\em et~al.}, ``{The upgrade
  of the ALICE TPC with GEMs and continuous readout}'',
  \href{https://doi.org/10.1088/1748-0221/16/03/P03022}{{\em JINST} {\bfseries
  16} (2021) P03022}, \href{https://arxiv.org/abs/2012.09518}{{\ttfamily
  arXiv:2012.09518 [physics.ins-det]}}.

\bibitem{ALICE:2025tba}
{\bfseries ALICE} Collaboration, I.~J. Abualrob {\em et~al.}, ``{Time
  resolution of the ALICE Time-Of-Flight detector with the first Run 3 pp
  collisions at ${\sqrt{\textit{s}} = 13.6}$ TeV}''
  \href{https://arxiv.org/abs/2511.10311}{{\ttfamily arXiv:2511.10311
  [physics.ins-det]}}.

\bibitem{ALICE:2013cdo}
{\bfseries ALICE} Collaboration, B.~B. Abelev {\em et~al.}, ``{$K^0_S$ and
  $\Lambda$ production in Pb--Pb collisions at $\sqrt{s_{\rm NN}}$ = 2.76
  TeV}'', \href{https://doi.org/10.1103/PhysRevLett.111.222301}{{\em Phys. Rev.
  Lett.} {\bfseries 111} (2013) 222301},
  \href{https://arxiv.org/abs/1307.5530}{{\ttfamily arXiv:1307.5530
  [nucl-ex]}}.

\bibitem{Podolanski:1954dyc}
J.~Podolanski and R.~Armenteros, ``{III. Analysis of V-events}'',
  \href{https://doi.org/10.1080/14786440108520416}{{\em Phil. Mag.} {\bfseries
  45} (1954) 13--30}.

\bibitem{ALICE:2025cqy}
{\bfseries ALICE} Collaboration, I.~J. Abualrob {\em et~al.}, ``{Centrality
  dependence of strange particle production in Pb-Pb collisions at sNN=5.02
  TeV}'', \href{https://doi.org/10.1103/j21y-pqd6}{{\em Phys. Rev. C}
  {\bfseries 114} (2026) 024909},
  \href{https://arxiv.org/abs/2511.10360}{{\ttfamily arXiv:2511.10360
  [nucl-ex]}}.

\bibitem{YuanyuanWang:2023vor}
Y.~Wang, Y.~Wang, W.~Zhao, and H.~Song, ``{Exploring the partonic collectivity
  in small systems at energies available at the CERN Large Hadron Collider}'',
  \href{https://doi.org/10.1103/n22l-1gdb}{{\em Phys. Rev. C} {\bfseries 113}
  (2026) 024901}, \href{https://arxiv.org/abs/2401.00913}{{\ttfamily
  arXiv:2401.00913 [nucl-th]}}.

\bibitem{Tang:2023wcd}
S.-Y. Tang, L.~Zheng, X.-M. Zhang, and R.-Z. Wan, ``{Investigating the elliptic
  anisotropy of identified particles in p{\textendash}Pb collisions with a
  multi-phase transport model}'',
  \href{https://doi.org/10.1007/s41365-024-01387-4}{{\em Nucl. Sci. Tech.}
  {\bfseries 35} (2024) 32}, \href{https://arxiv.org/abs/2303.06577}{{\ttfamily
  arXiv:2303.06577 [hep-ph]}}.

\end{thebibliography}\endgroup
